\documentclass[onecolumn]{IEEEtran}%twoside, lettersize, onecolumn]{IEEEtran}

\usepackage[
    markup=default,
    authormarkup=none,
    final
]{changes}

\usepackage{amsmath, amssymb, tikz, enumitem, tabularx, bm, amsthm, mathabx, import, subfiles, subfigure, mathtools, tikz-cd, xfrac, dsfont, authblk, xr, lineno, xcolor}

\usepackage{marginnote}
\newif\ifshowlabels         % master switch

\showlabelsfalse            % ← final

\newcommand{\revlbl}[1]{%
  \ifshowlabels
    \marginnote{\footnotesize\color{purple}{\textup{[#1]}}}%
  \fi}
\newtheorem{proposition}{Proposition}
\newtheorem{lemma}[proposition]{Lemma}

\newtheorem{theorem}[proposition]{Theorem}

\newtheorem{corollary}[proposition]{Corollary}

\theoremstyle{definition}
\newtheorem{definition}[proposition]{Definition}
\newtheorem{example}[proposition]{Example}
\newtheorem{remark}[proposition]{Remark}

\newcommand{\Z}{\ensuremath{\mathbb{Z}}}% integers
\newcommand{\N}{\ensuremath{\mathbb{N}}}% natural numbers
\newcommand{\R}{\ensuremath{\mathbb{R}}}% real numbers
\newcommand{\inv}{\ensuremath{^{-1}}}

\newcommand{\cat}[1]{\ensuremath{\bm{\mathrm{#1}}}}

\newcommand{\phid}{\ifmmode \Phi\,\mathrm{ID} \else $\Phi\,\mathrm{ID}$ \fi}

\newcommand\indep{\perp\!\!\!\perp}

\DeclareMathOperator{\Rep}{Rep}
\DeclareMathOperator{\TC}{TC}
\DeclareMathOperator{\DTC}{DTC}
\DeclareMathOperator*{\Star}{\bigast}

\renewcommand{\phi}{\varphi}
\renewcommand{\epsilon}{\varepsilon}

\usetikzlibrary{decorations.markings, shapes, positioning}

\usepackage{graphicx} % Required for inserting images

\title{A Logarithmic Decomposition and a Signed Measure Space for Entropy}

\author[1,2]{Keenan J. A. Down}
\author[3,4]{Pedro A. M. Mediano}

\affil[1]{\textit{Department of Psychology, Queen Mary, University of London}}
\affil[2]{\textit{Department of Psychology, University of Cambridge}}
\affil[3]{\textit{Department of Computing, Imperial College London}}
\affil[4]{\textit{Division of Psychology and Language Sciences, University College London}\thanks{This paper was presented in part at the IEEE International Symposium on Information Theory 2023 (ISIT).}}

\date{August 2023}

\usepackage{subfiles}

\begin{document}

\maketitle

\begin{abstract}
The Shannon entropy of a random variable has much behaviour analogous to a signed measure. Previous work has explored this connection by defining a signed measure on abstract sets, which are taken to represent the information that different random variables contain. This construction is sufficient to derive many measure-theoretical counterparts to information quantities such as the mutual information (the intersection of sets), the joint entropy (the union of sets), and the conditional entropy (the difference of sets). Here we provide concrete characterisations of these abstract sets and a corresponding signed measure \revlbl{R1.1}\added{by extending the approach used by Yeung to all possible outcomes in an outcome space $\Omega$}, and in doing so we demonstrate that there exists a much finer decomposition with intuitive properties which we call the \textit{logarithmic decomposition (LD)}. We show that this signed measure space has the useful property that its \textit{logarithmic atoms} are easily characterised with negative or positive entropy\added{, depending only on their structure}, while also being consistent with Yeung's I-measure. We present the usability of our approach by re-examining the G\'acs-K\"orner common information\revlbl{R2.1} \added{ and the minimum sufficient statistic} from this new geometric perspective and characterising it in terms of our logarithmic atoms -- a property we call \textit{logarithmic decomposability}. We present possible extensions of this construction to continuous probability distributions before discussing implications for quality-led information theory. \added{As a motivating example}, we apply our new decomposition to  the Dyadic and Triadic systems of James and Crutchfield and show that, in contrast to the I-measure alone, our decomposition is able to qualitatively distinguish between them.\revlbl{R1.1} \added{Previously it has been believed that classical measures are unable to distinguish the two; as our decomposition is fundamentally classical, we demonstrate this to be false.}
\end{abstract}


\section{Introduction}

\subsection{Background}

It was shown by Yeung in 1991 that for all first-order information-theoretical quantities derived from the classical Shannon entropy on a collection of random variables $X_1,\ldots, X_r$, there is a corresponding set in a $\sigma$-algebra $\mathcal{F}$, and, moreover, that for any set in the $\sigma$-algebra there exists a corresponding measure of information \cite{yeung1991new}. Yeung's $I$-measure is a signed measure on this $\sigma$-algebra and can be constructed by symbolic substitution on classical information quantities. This correspondence between abstract sets and information quantities, built upon earlier work by Hu Kuo Ting \cite{ting1962amount}, offers a firm foundation for the measure-theoretical perspective of Shannon entropy, but remains relatively coarse. For example, when constructing the G\'acs-K\"orner common information variable $C(X_1;\ldots, X_r)$ for a collection of variables $X_1,\ldots, X_r$ \cite{gacs1973common}, the $I$-measure provides no strong insight into where this variable comes from. In the same work, G\'acs and K\"orner went so far as to present their original aim as `to show that common information has nothing to do with mutual information'. A finer measure might offer some resolving ability to see which pieces of the information should be contained in the common information variable and which should not.

Another classic example of the coarseness of the $I$-measure is that there exist systems which are, by construction, qualitatively distinct, yet cannot be discerned using the measure alone. To see this, one might consider the Dyadic and Triadic systems highlighted by James and Crutchfield \cite{james2017multivariate} (see section \ref{SECTION_TriadicDyadic}). These two systems, despite being qualitatively different, cannot be discerned using the $I$-measure alone, and their entropies, conditional entropies and co-informations are completely identical under the measure.

Yeung's correspondence draws a formal relationship between various operations on random variables and operations on sets. Given a collection of random variables $X_1,\ldots, X_r$, the $\sigma$-algebra as constructed by Yeung is generated by the unions, intersections, and complements of various \textit{set variables} $\tilde{X}_1, \ldots, \tilde{X}_r$ \cite{yeung1991new}, which can be taken symbolically to represent ``spaces'' of information; sets which can be thought of as containing the information held by a variable. The construction as given by Yeung is entirely symbolic and does not attempt to characterise the constituent elements of these spaces.

This connection between information theory and measure theory is mechanically useable and consistent, but the contents of the spaces $\tilde{X}_1, \ldots, \tilde{X}_r$ remains mysterious. Indeed, the set-theoretic structure in this case is built entirely using the already-known information theoretic structure, so this perspective contributes little to the intuition of random variables as \textit{sets of information}. In principle, the construction is completely symbolic, and reasoning in terms of sets seems to add little additional intuition.

Under the given correspondence, Yeung showed we are justified in making a substitution of symbols:
\begin{align}
    X_1, X_2, \ldots, X_r \quad &\longleftrightarrow \quad \tilde{X}_1, \tilde{X}_2, \ldots, \tilde{X}_r \notag\\
    H(X) \quad &\longleftrightarrow \quad \mu_Y(\tilde{X}) \notag\\
    H(X \,|\, Y) \quad &\longleftrightarrow \quad \mu_Y(\tilde{X} \, \setminus \, \tilde{Y}) \\
    H(X, Y) \quad &\longleftrightarrow \quad \mu_Y(\tilde{X} \cup \tilde{Y}) \notag\\
    I(X; Y) \quad &\longleftrightarrow \quad \mu_Y(\tilde{X} \cap \tilde{Y}), \notag
\end{align}
where we have taken $I(X,Y)$ to represent the mutual information between $X$ and $Y$, and we write $\mu_Y$ to represent the $I$-measure of Yeung.

Decomposing these information spaces would be of great interest across multiple domains. What kind of information is transmitted across a network of neurons and with what qualitative structure does it possess \cite{luppi2022synergistic, ince2017statistical, gelens2024distributed}? How is information manipulated, digested and represented in a machine learning model (the problem of developing explainable AI) \cite{angelov2021explainable, burkart2021survey, proca2024synergistic}? How can we disentangle the complex interplay between confounding variables, such as gender and job acceptance, or race and arrest rate \cite{mehrabi2021survey}? Understanding the composition of information itself at various structural scales (at least, beyond symbolic substitution) might play a key role in providing new avenues for answering these kinds of questions. Such decompositions might also allow us to understand how coding properties of mutual information and co-information relate to the variables that generate them, despite not being generally representable by a variable \cite{gacs1973common}. \revlbl{R1.1}\added{The current perspective from the literature is that partial information decomposition (PID) is required in order to distinguish between these two systems. We will show in this work that this is, in fact, not the case, and that by refining the $I$-measure of Yeung, the structural differences between these systems can be revealed.}

\subsection{\added{Relation to Partial Information Decomposition (PID) and other decomposition methods}}
\revlbl{E.2}\added{Partial Information Decomposition (PID) is a method initially introduced by Williams and Beer in 2010 under the premise that, given some set of `source variables' $X_1,\ldots, X_n$ and a target variable $T$ about which one wishes to obtain information, the mutual information $I(X_1,\ldots,X_n; T)$ can be decomposed into \textbf{redundant}, \textbf{unique} and \textbf{synergistic} components \cite{williams2010nonnegative}. Having found the initially proposed definition of redundancy to be unsatisfactory, many other alternative versions of the PID methodology have since been proposed \cite{bertschinger2014quantifying, quax2017quantifying, james2018unique, griffith2014intersection, griffith2014quantifying, griffith2015quantifying, harder2013bivariate, ince2017measuring, finn2018pointwise, kolchinsky2022novel, barrett2015exploration}.}

\added{The logarithmic decomposition presented here does not, in and of itself, correspond to a method of partial information decomposition as normally formulated. However, it \textit{does} imply certain ways of thinking about redundancy and synergy \cite{down2025algebraic}. Most notably, in section \ref{SECTION_TriadicDyadic}, we propose a subset $R_2$ (labelled so as to imply redundancy), which is sufficient to distinguish between the dyadic and triadic systems. While this set \textit{intuitively looks} like a generalisation of mutual information to three variables, it fails to be a natural measure for redundancy in the PID sense, as, under utmost generality, $R_2$ can also have negative measure (though this appears to be quite rare). Despite this, we believe LD might, with some careful application, allow either for the construction of a `classical' PID (stronger than MMI), or be a useful tool for showing that partial information decomposition cannot obey relevant chain rules, potentially pointing at a fundamental flaw in the method. We leave this exploration to future work.}

\revlbl{E.1}\added{A similar decomposition, inspired by the logarithmic decomposition presented here, has also been proposed by Li using arrival times in Poisson processes \cite{li2023poisson}. For more detail on the relationship we refer the reader to appendix \ref{SECTION_Simplex}.}

\subsection{Main contributions}

In the present work we describe these information spaces in greater detail than has previously been seen. Given a collection of random variables $X_1,\ldots, X_r$ on a joint outcome space $\Omega$, we present a theoretically maximal refinement of the corresponding $\sigma$-algebra, which we label $\Delta \Omega$. We demonstrate in which sense it is maximal in Appendix~\ref{SECTION_Simplex}. Given this refined space $\Delta\Omega$, we will then construct a signed measure we call the \textbf{interior loss}, $\mu$, which shall represent the entropy content of the measurable sets in the space. In doing so, we decompose the $\sigma$-algebra of Yeung \cite{yeung1991new} into many fine pieces we call \textbf{logarithmic atoms}, whose contribution to the entropy is particularly easy to characterise with surprising parity properties, in a process and paradigm we have labelled \textbf{logarithmic decomposition}. This decomposition might be viewed as a natural extension of an earlier construction by Campbell \cite{campbell1965entropy}, whose constructed measure dealt exclusively with equiprobable outcomes on \added{independent} variables.

From this new perspective, the abstract information spaces $\tilde{X}_1,\ldots, \tilde{X}_r$ are now fully realised. Using this decomposition, they can now be seen to contain multiple atoms of information, each with a single qualitative interpretation which makes them particularly pleasant to characterise. These atoms are in bijection with subsets of the outcome space $\Omega$ with singlets and the empty set removed, and whether or not a given random variable \textit{has knowledge} of a given atom is also straightforward to characterise. That is, as a set, it is quite straightforward to determine the set-theoretic composition of the information space $\tilde{X}$.

In sections \ref{SECTION_Refinement} and \ref{SECTION_Construction} we construct the signed measure space $(\Delta \Omega, \mu)$ by describing a set of \textit{atoms} of information. Subsets of this space will form the elements of the abstract information spaces $\tilde{X}_i$, which we later refer to as $\Delta X_i$. We also prove many useful results on the measures of individual atoms. For example, we demonstrate that for any given atom $b$, the sign of the contribution $\mu(b)$ is fixed by its structure -- a property lost at coarser resolutions. \revlbl{R1.1, R1.2}\added{In particular, the $I$-measure, now viewed as a collection of these refined atoms, does not allow for knowledge of the sign of its components ahead of time. Under this refinement, the sign of every contribution is accounted for, even without any knowledge of the underlying probabilities in the system. In particular, this allows for the structural investigation of various systems - in \cite{down2025algebraic}, a sequel to this work, we use this unique property to study synergistic information, showing that, allowing probabilities to vary freely, the XOR gate is the only  system of three variables $X,Y, f(X,Y)$ with purely synergistic (purely negative co-information) behaviour.}

In section \ref{SECTION_Quantities} we \added{ show that our measure both refines and is consistent} with the $I$-measure \cite{yeung1991new}. We characterise the entropy of a variable $H(X)$ as the total measure of all atoms in its information space, $\mu(\Delta X)$, and we show that the mutual information also has a representation as $\mu(\Delta X \cap \Delta Y)$. Additionally, we recover natural representations for the common information of G\'acs and K\"orner \cite{gacs1973common} \revlbl{R2.1, R2.2}\added{and the minimum sufficient statistic}. We give a description of these \textit{logarithmically decomposable} quantities; quantities which have a set-theoretic representation under our decomposition.

In sections \ref{SECTION_Inclusions} and \ref{SECTION_Continuous} we develop the theory to explain how information representations change when refining the outcome space $\Omega \to \Omega'$ and how this can be applied to study continuous variables. In doing so, we recover the limiting density of discrete points of Jaynes \cite{jaynes1957information, jaynes1968prior}. Using this, we give a novel set-theoretic perspective on why, under refinements, mutual information is often bounded while entropy is not.

As a final demonstration of the utility of this decomposition \added{and the main result of this paper}, we apply our methods in section \ref{SECTION_TriadicDyadic} to the Dyadic and Triadic systems of James and Crutchfield \cite{james2017multivariate}, where we shall see it has the ability to discern between these two systems -- an improvement over the classical $I$-measure. \revlbl{R1.1, R1.2}\added{In doing so, we show that the prevailing belief in the literature-- that these two systems cannot be distinguished without extensions to classical information theory (such as Partial Information Decomposition-- see \cite{williams2010nonnegative,bertschinger2014quantifying,ince2017measuring,kolchinsky2022novel} for example) is, in fact, false.}

\revlbl{R1.2}\added{A sequel to the results of this paper is available, where further interrogation of the structure of this decomposition is performed. Therein, we make use of this decomposition to explore other problems relevant to Partial Information Decomposition \cite{down2025algebraic}. In particular, the decomposition presented here has been applied to bound co-information, with the powerful result that the XOR gate is the only system $X, Y, f(X,Y)$ which has purely synergistic behaviour.}

The proofs of all results, where not insightful, are included in the appendix.

\ifSubfilesClassLoaded{\bibliographystyle{plain} \bibliography{main} }{}
\end{document}

\section{An explicit definition for abstract information spaces}
\label{SECTION_Refinement}

Let $\Omega$ be a discrete sample space. When considering a collection of variables $X_1,\ldots X_r$, we require $\Omega$ to be at least as fine as the joint outcome space for $X_1X_2\ldots X_r$. Let $\mathcal{F}$ be the natural $\sigma$-algebra generated by all combinations of outcomes on each variable and let $P$ be a probability measure on $\Omega$. We shall use the probability space $(\Omega, \mathcal{F}, P)$ to define a corresponding space for information.

\begin{definition}
Let $(\Omega, \mathcal{F}, P)$ be a probability space as above. Then we define the \textbf{content} of $\Omega$ to be the simplicial complex on all outcomes $\omega \in \Omega$, with the vertices removed:
\begin{equation}
\Delta\Omega = \bigcup_{k = 2}^N \Omega_k \cong \mathcal{P}(\Omega) \setminus \left( \{\{\omega\}: \omega \in \Omega\} \cup \{\varnothing\} \right)
\end{equation}
where $\Omega_k$ is the set of subsets $S \subseteq \Omega$ with $|S| = k$ and $N = |\Omega|$. For a collection of $n$ outcomes $\omega_1,\ldots, \omega_n$, we label the corresponding simplex as $b_{\omega_1\omega_2 \ldots \omega_n}$ or simply $\omega_1\omega_2\ldots \omega_n \in \Omega^n$ for ease of notation. Viewing $\Delta\Omega$ geometrically as a simplex, this element will correspond to a face, volume, or edge on a simplex without its boundaries.

For consistency we have opted to exclude single outcomes (vertices on the simplex) and the empty set $\varnothing$. We will see later that these parts of the space do not contribute to the entropy and are not necessary for the construction of the measure space.
\end{definition}

\begin{example}
Consider a space of outcomes $\Omega = \{1, 2, 3, 4\}$. The content space consists of the following elements
\begin{align}
\begin{split}
\Delta\Omega = \{& b_{12}, b_{13}, b_{14},\\
& b_{23}, b_{24}, b_{34}, \\
& b_{123}, b_{124}, b_{134}, b_{234}, \\
& b_{1234}\}
\end{split}
\end{align}

Subsets of this space will correspond in the sequel to representations of different information quantities. For example, the subset $\{b_{12}, b_{14}, b_{24}, b_{124}\}$ as in figure \ref{FIGURE_124Triangle}. We will see later that, despite being a measurable quantity, this set cannot be represented by a variable.

\begin{figure}[h]
\centering
\begin{tikzpicture}[line join = round, line cap = round]

\coordinate [label=above:4] (4) at (0,{sqrt(2)},0);
\coordinate [label=left:3] (3) at ({-.5*sqrt(3)},0,-.5);
\coordinate [label=below:2] (2) at (0,0,1);
\coordinate [label=right:1] (1) at ({.4*sqrt(3)},0,-.5);

\begin{scope}[decoration={markings,mark=at position 0.5 with {\arrow{to}}}]
\draw[dotted] (1)--(3);
\draw[very thick, gray, fill=lightgray,fill opacity=.5] (2)--(1)--(4)--cycle;
\draw[dotted] (3)--(2);
\draw[dotted] (3)--(4);
\end{scope}

\end{tikzpicture}

\caption{The highlighted triangle along with its boundary corresponds to the subset $\{b_{12}, b_{14}, b_{24}, b_{124} \}$.}
\label{FIGURE_124Triangle}
\end{figure}
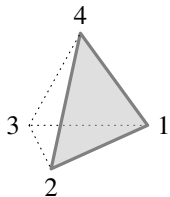

\end{example}

In the theory of lattices and order, an \textbf{atom} is a minimal nonzero element. For example, when ordering a Venn diagram under inclusion, the smallest regions of the diagram are the smallest nonzero structures under the order -- these pieces (the atoms) form additive building blocks for all other objects. For that reason, we often refer here to the individual pieces of the content space $\Delta \Omega$ as \textbf{atoms}, as is commonplace in the theory of information decomposition \cite{shannon1953lattice, williams2010nonnegative, bell2003co}. \revlbl{R1.1}\added{In the framework of the $I$-measure, atoms correspond to all conditional entropies and conditional mutual informations. Each separate and indivisible region of an $I$-diagram (the set-theoretic representation) corresponds to a different atom \cite{yeung1991new}. Under our framework, atoms are built by considering the entire outcome space and separating distinct events within variables. In doing so, we split the atoms of the $I$-measure down further into even smaller pieces, producing the representation defined here.}

\added{
\begin{remark}
\label{REMARK_Measure_everything}
An alternative way to view this decomposition is via the $I$-measure itself, but deliberately incorporating maximal data on the outcome space. For example, if one wishes to study the interactions between the variables $X_1,\ldots, X_n$, one can also define an additional collection of variables $T_\omega$ for $\omega \in \Omega$ where each $T_\omega$ is the indicator variable for $\omega \in \Omega$. The atom $b_S$ corresponding to a collection of outcomes $S \subseteq \Omega$ will then lie at the intersection $I(T_\omega; \, \omega\in S)$. In this system, all conditional entropies $H(T_\omega | T_{\omega'}: \omega' \in \Omega)$ are precisely zero, and hence are discarded.
\end{remark}
}

\begin{remark}
It is no accident that we have used the notation $\Delta\Omega$ to represent the set of atoms in our construction. We shall see later in section \ref{SECTION_Quantities} that individual atoms correspond at an operational level to a given variable's ability to distinguish between outcomes. That is to say, we shall see that when the variable captures information about a \textit{change} between two or more outcomes, that atom becomes part of the information space corresponding to $X$. This will be concretised in section \ref{SECTION_Quantities}.
\end{remark}

In this section we have treated the discrete case. For an extension into the continuous case, it is necessary to consider successive refinements of discrete spaces. We explore this in sections \ref{SECTION_Inclusions} and \ref{SECTION_Continuous}.

In the next section, we construct the measure $\mu$ to accompany this space. Doing so will complete the construction of the refined signed measure space $(\Delta\Omega, \mu)$. \revlbl{R1.1}\added{We shall see that by refining these spaces of information (that is, further decomposing  their constituent atoms) we recover useful properties that are lost at the coarser scale of the $I$-measure.}

\ifSubfilesClassLoaded{\bibliographystyle{plain} \bibliography{main} }{}
\end{document}

\section{Construction of a signed measure}
\label{SECTION_Construction}

Having endowed $\Delta\Omega$ with a geometric interpretation, we would like to equip it now with a signed measure. Such a space will provide a qualitative and quantitative language for information; subsets in the measure space representing a \textbf{quality}, and the measure of those subsets representing the \textbf{quantity}. With this completed measure space in hand, we will be able to proceed with a refined description of the \textbf{information spaces} $\tilde{X}_j$ of random variables $X_j$ over the outcome space $\Omega$, which is to be desired to fully flesh out the correspondence between random variables and set-variables \cite{ting1962amount, yeung1991new}. In order to construct these spaces we will need to develop the language to handle the information encoded by any event defined on the outcome space $\Omega$, and we shall see that the space $\Delta \Omega$ provides a sound underlying set for such quantities.

We will build our measure $\mu$ on finite collections of atoms by considering the notion of entropy loss, an alternative perspective from which it is possible to re-derive the classical Shannon information measure. Baez, Fritz and Leinster showed in \cite{baez2011characterization} that rather than considering a direct formula for entropy, one could measure the entropy of a random variable $X$ by considering the \textit{loss in entropy} under a mapping $f: X \to 1$; a morphism to the trivial partition. Similarly, any mapping $f: X \to Y$ will be associated with an entropy loss. Entropy loss \added{has} properties which absolute entropy does not possess. For example, the authors demonstrated in the same work that entropy loss is homogeneous \cite{baez2011characterization}, and this property will be useful when building our decomposition.

In this work we refer to this idea as the \textbf{total entropy loss} or \textbf{loss}, $L$. From this we will then construct the measure $\mu$ of our signed measure space using a M\"obius inversion. For geometric reasons, we occasionally refer to the measure $\mu$ as \textbf{interior loss}.

\revlbl{R1.1}\added{We note that using the entropy loss appears to be a more convenient approach for describing this decomposition, rather than the approach given in remark \ref{REMARK_Measure_everything}. In particular, using the entropy loss, we sidestep the need to construct variables with which to describe the measure.}

The final signed measure space shall then consist of the signed measure $\mu$ and the space $\Delta\Omega$. We will see that, geometrically, the total entropy loss $L$ will measure entire simplices inside of $\Delta \Omega$ with their boundaries, while $\mu$ will measure the interiors of these simplices alone - boundaries not included - hence the name \textbf{interior loss}.

Using the perspective of entropy loss, we shall say that a variable will \textit{lose} entropy when boundaries between events are deleted \cite{baez2011characterization}, so that two or more events are merged into a single event. More concretely, let $X$ be a random variable corresponding to a partition $\bm{Q}_X = \{Q_1,\ldots, Q_t\}$ of the outcome space $\Omega$ where $P(Q_i) = \sum_{\omega \in Q_i} P(\omega)$ for finite $\Omega$, and $\sum_{k = 1}^t P(Q_k) = 1$. If we create a new random variable $X'$ by merging two of the events given by parts $Q_1$ and $Q_2$ so that $\bm{Q}_{X'} = \{Q_1 \cup Q_2, Q_3, \ldots, Q_t\}$ becomes the new partition, then the new variable $X'$ will have a reduced entropy. In particular, note that if we remove all boundaries and merge all events in a variable into a single outcome, then the corresponding entropy loss will be the total entropy of $X$, $H(X)$.

\begin{definition}
\label{DEFINITIONTotalLoss}
Let $X$ be a random variable with corresponding partition $\bm{Q}_X = \{Q_1, \ldots, Q_t\}$, and let $X'$ be the random variable with corresponding partition
\begin{equation}
\bm{Q}_{X'} = \left\{ \bigcup_{a \in A} Q_a \right\} \cup \{Q_b: b \notin A\},
\end{equation}
where $A$ is a subset of $n$ events which we intend to merge, so that these events correspond to a single event in the new variable. In particular, $\bm{Q}_{X'}$ is given by taking $\bm{Q}_X$ with all parts indexed in $A$ merged together. We then define the corresponding \textbf{total entropy loss}
\begin{equation}
\label{EqnEntropyChange}
L(A) = H(X) - H(X').
\end{equation}
We may simplify the notation somewhat and write $L(p_1,\ldots, p_n)$, where the $p_i = P(Q_i)$ are the probabilities associated with each part or event in the set $A$. Doing this also emphasises that $L$ can also be viewed as a function on $[0,1]^n$. Expanding the above expression we find
\begin{align}
\label{EQN_LossFormula}
\begin{split}
L(p_1,\ldots, p_n) =&\, H(X) - H(X') \\
= &\, p_1 \log \left(\frac{1}{p_1}\right) + \cdots + p_n \log \left(\frac{1}{p_n}\right) \\
 &\, - (p_1 + \cdots + p_n) \log \left( \frac{1}{p_1 + \cdots + p_n} \right) \\
= &\, \log \left[ \frac{(p_1 + \cdots + p_n)^{(p_1 + \cdots + p_n)}}{p_1^{p_1} \ldots p_n^{p_n}} \right] .
\end{split}
\end{align}
\end{definition}

\begin{remark}
This definition is equivalent to considering the entropy loss on a variable $X$ after the mapping
\begin{equation}
f: X \to X'
\end{equation}
\begin{equation}
f(x) = \begin{cases}x & x\notin A \\
1 & x \in A \end{cases}
\end{equation}
where $1$ denotes some symbol not already in the alphabet of $X$.
\end{remark}

It is worth briefly remarking that $L(A) \geq 0$ given any collection of parts $A$. Moreover, using equation \ref{DEFINITIONTotalLoss}, it is immediately clear that for a random variable $X$ with events of associated probabilities $p_1,\ldots, p_n$ with $\sum p_i = 1$, we must have
\begin{equation}
\label{EQUATION_EntropyIsLoss}
    H(X) = L(p_1,\ldots, p_n).
\end{equation}
Trivially we also see that $L(p) = 0$ for any single $p \in [0,1]$, as merging one event with itself does not result in a loss of entropy. Note that in the case that the $p_i$ do \textit{not} sum to one the property that $L(p_1,\ldots, p_n) = H(p_1,\ldots, p_n)$ does not hold; the expected log surprisal will no longer be equal to the loss. We shall see shortly that this behaviour offers some additional algebraic properties that the classical measure does not possess. In addition to this, we shall demonstrate in subsection \ref{SUBSECTION_EntropyLossProperties} that the behaviour of entropy loss endows our construction with a new perspective to the original axioms on $H(X)$ given by Shannon in his original paper \cite{shannon1948mathematical}.

Loss alone is not sufficient to construct a refined signed measure space for information, as it is only additive through the composition of morphisms or across disjoint systems. To account for this, we now supplement the definition of the total loss with a M\"obius inversion to construct an additive measure $\mu$. This $\mu$, which we call the \textbf{interior loss}, will be the measure attached to our refined measure space for Shannon entropy.

For maximum strength in our construction, we will now treat $\Omega$ as a partition of singletons $\omega_i \in \Omega$, as this is is sufficiently rich in structure to describe all variables defined on this space.

\begin{remark}
As our goal is to construct a measure space, it will often be convenient to allow the loss $L$ (and the measure $\mu$) to be defined on both outcomes and on probabilities. For this purpose we shall also allow ourselves to use outcomes as function arguments, where we implicitly take
\begin{equation}
L(\omega_1,\ldots, \omega_n) := L(P(\omega_1), \ldots, P(\omega_n)).
\end{equation}
Similarly, given a set $S = \{\omega_1,\ldots, \omega_n\}$, we allow ourselves to write
\begin{equation}
L(S) = L(\omega_1,\ldots, \omega_n) = L(P(\omega_1),\ldots, P(\omega_n)).
\end{equation}
Note that we will often have arguments $L(p_1,\ldots, p_n)$ where the $p_i$ do not sum to one. In fact, the theory that follows appears to be completely agnostic of the requirement that the \textit{probabilities} sum to one.
\end{remark}

\begin{definition}
\label{DEFINITION_InteriorLoss}
We will define the \textbf{interior loss} $\mu(\omega_1,\ldots, \omega_n)$ recursively on the number of outcomes which are being merged. For $n=1$ let $\mu(\omega) = 0$. For $n\geq 1$ we define $\mu$ by
\begin{equation}
\mu(\omega_1,\ldots, \omega_n) = L(\omega_1,\ldots, \omega_n) - \sum_{\substack{S \subset \{\omega_1,\ldots, \omega_n\} \\ |S| \leq n-1}} \mu(S).
\end{equation}
This construction corresponds to a M\"obius inversion on the lattice of subsets of outcomes $\mathcal{P}(\Omega)$, where the partial order is given by inclusion. Again, as with the total loss, we will often abuse this notation and write $\mu(p_1,\ldots, p_n)$ where the probabilities reflect individual outcomes or regions in the partition.
\end{definition}

In the geometric framework of the previous section, we can think of $\mu$ as measuring entropies in interior regions of the simplex $\Delta\Omega$. That is to say, $\mu$ can be thought of as measuring faces, edges, or volumes without their boundaries, while the total loss $L$ can be thought of as measuring simplices with their boundaries included. The M\"obius inversion on the loss enables us to assign entropy contributions to the interiors of these simplices.

Restated, the purpose of the M\"obius inversion is to reclaim additivity: it converts the not-always-additive measure $L$ to the additive measure $\mu$ (as is necessary for the set-theoretic perspective). We will later see that it is not always possible to express mutual information using a positive sum of losses alone; one requires the measure $\mu$ to recover it in general. Its use here should be further justified by theorem \ref{AlternatingDerivatives}, which we prove in the next subsection.

\revlbl{R2.3}
\begin{example}
\added{Consider the following system of four outcomes $\Omega = \{1,2,3,4\}$ with probabilities $0.16, 0.34, 0.23$ and $0.27$, respectively.}

\begin{figure}[ht]
\centering
    \scalebox{0.88}{
    \begin{tikzpicture}[scale=0.88]
    %%% XYZ with 1, 2, 3, 4
    \draw (0,0) -- (2, 0);
    \draw (0,0) -- (0, -2);
    \draw (0, -2) -- (2, -2);
    \draw (2, -2) -- (2, 0);
    \draw (0, -1) -- (2, -1);
    \draw (1, 0) -- (1, -2);
    \node at (0.5, -0.5) {$(1)$};
    \node at (1.5, -0.5) {$(2)$};
    \node at (0.5, -1.5) {$(3)$};
    \node at (1.5, -1.5) {$(4)$};
    \node at (1, -2.35) {$\Omega$};

    %%% XYZ in probabilities
    \draw (3,0) -- (5, 0);
    \draw (3,0) -- (3, -2);
    \draw (3, -2) -- (5, -2);
    \draw (5, -2) -- (5, 0);
    \draw (3, -1) -- (5, -1);
    \draw (4, 0) -- (4, -2);
    \node at (3.5, -0.5) {$0.16$};
    \node at (4.5, -0.5) {$0.34$};
    \node at (3.5, -1.5) {$0.23$};
    \node at (4.5, -1.5) {$0.27$};
    \node at (4, -2.35) {$\Omega$};

    %%% X1
    \begin{scope}[shift = {(3,1.5)}]
    \draw[->] (2.25, -1.25) -- (2.75,-0.75);
    \draw[->] (5.25, -0.75) -- (5.75,-1.25);
    \draw (3,0) -- (5, 0);
    \draw (3,0) -- (3, -2);
    \draw (3, -2) -- (5, -2);
    \draw (5, -2) -- (5, 0);
    \draw (3, -1) -- (5, -1);
    \draw (4, -1) -- (4, -2);
    \node at (4, -0.5) {$0.50$};
    \node at (3.5, -1.5) {$0.23$};
    \node at (4.5, -1.5) {$0.27$};
    \node at (4, -2.35) {$X_1$};
    \end{scope}

    %%% X2
    \begin{scope}[shift = {(3,-1.5)}]
    \draw[->] (2.25, -0.75) -- (2.75,-1.25);
    \draw[->] (5.25, -1.25) -- (5.75,-0.75);
    \draw (3,0) -- (5, 0);
    \draw (3,0) -- (3, -2);
    \draw (3, -2) -- (5, -2);
    \draw (5, -2) -- (5, 0);
    \draw (4, -1) -- (5, -1);
    \draw (4, 0) -- (4, -2);
    \node at (3.5, -1) {$0.39$};
    \node at (4.5, -0.5) {$0.34$};
    \node at (4.5, -1.5) {$0.27$};
    \node at (4, -2.35) {$X_2$};
    \end{scope}

    %%% X
    \draw (9,0) -- (11, 0);
    \draw (9,0) -- (9, -2);
    \draw (9, -2) -- (11, -2);
    \draw (11, -2) -- (11, 0);
    \draw (11, -1) -- (10, -1) -- (10, -2);
    \node at (9.5, -0.5) {$0.73$};
    \node at (10.5, -1.5) {$0.27$};
    \node at (10, -2.35) {$X$};

    \end{tikzpicture}
    }
\caption{\added{Treating $\Omega$ itself as a random variable (possibly as the joint of all variables being considered, $\Omega$ has entropy $H(\Omega) = 1.95$ bits. After merging the three outcomes, $X$ has entropy $H(X) = 0.84$ bits, having lost more than a bit of entropy. There are different intermediate ways of merging outcomes which lose less, but different entropies en route to $X$.}}
\label{FIG_two_splitting_paths}
\end{figure}

\added{
In this setting, we can treat $\Omega$ itself as a variable in order to compute the entropy losses associated with different subsets of $\Omega$. With the probabilities as given, the maximum possible entropy of a variable defined on $\Omega$ is $H(\Omega) = 1.95$ bits. If we merge three of these outcomes-- 1, 2 and 3, say, then we lose $L(\omega_1, \omega_2, \omega_3) = L(1,2,3) = L(0.16, 0.34, 0.23)$ bits of entropy, where
\begin{equation}
L(0.16, 0.34, 0.23) = H(\Omega) - H(X) = 1.95 - 0.84 = 1.11 \text{ bits}.
\end{equation}
}
\added{
The total loss $L(1,2,3)$ represents the entropy lost when grouping all three outcomes together. However, doing so presupposes that any subset of $\{1,2,3\}$ is \textit{also} grouped together. Here, for example, the entropy that $X_1$ has lost is only $L(1,2) = 0.45$ bits, and $X_2$ has lost $L(1,3) = 0.38$ bits of entropy. For that reason, the entropy lost in $L(1,2,3)$ is coming \textit{more} from the merger of outcomes $1$ and $2$, rather than $1$ and $3$ (the same can be done with $2$ and $3$, where the attributed loss would be $0.55$ bits).
}
\added{
When we group together outcomes $1, 2$ and $3$ to create the variable $X$, we implicitly do all of the different pairwise groupings. However, $L(1,2,3)$ is \textit{not} equal to $L(1,2) + L(1,3) + L(2,3)$; we would be over-counting the entropy loss in this case (later made explicit in Theorem \ref{AlternatingDerivatives}).
}
\added{
To account for this \textit{over-estimation}, we can attribute some entropy to the trio $1,2,3$ itself, over and above the pairwise interactions. In order to compute the true entropy loss $\Omega \to X$, we can take the entropy $L(1,2,3)$ and subtract the contributions of $L(1,2)$, $L(1,3)$ and $L(2,3)$. Doing so, we obtain the \textit{interior loss} $\mu$; the entropy intrinsically connected to the trio $1,2,3$, over and above its constituent pairs.
\begin{equation}
    \mu(1,2,3) = L(1,2,3) - L(1,2)-L(1,3)-L(2,3) = -0.28 \text{ bits}.
\end{equation}}
\added{
From this perspective, $\Omega \to X_1$ loses $L(1,2) = \mu(1,2)$ bits of entropy, $\Omega \to X_2$ loses $\mu(1,3)$ bits of entropy, and $\Omega \to X$ loses $\mu(1,2) + \mu(1,3) +\mu(2,3) + \mu(1,2,3)$ bits of entropy.}
\end{example}

\begin{remark}
\label{REMARK_InclusionExclusionIdentities}
The total loss can be expressed as a sum of the interior losses by virtue of their construction:
\begin{equation}
L(\omega_1,\ldots, \omega_n) = \sum_{S \subseteq \{\omega_1,\ldots, \omega_n\}} \mu(S),
\end{equation}
and hence the interior loss function can also be expressed in terms of the loss function by virtue of the inclusion-exclusion principle \cite{stanley1986enumerative}:
\begin{equation}
\label{InclusionExclusion}
\mu(\omega_1,\ldots, \omega_n) = \sum_{S \subseteq \{\omega_1,\ldots, \omega_n\}} (-1)^{n-|S|} L(S).
\end{equation}
The interior loss corresponds to the M\"obius inversion of the total loss on the partially ordered set defined by containment of simplices.
\end{remark}

The expression in equation \ref{EQN_LossFormula} appears to imply that the functions $L$ and $\mu$ can both be extended to domains where the probabilities $p_i$ are greater than one, or do not sum to one, and as it turns out, all of the results in this paper (aside from equation \ref{EQUATION_EntropyIsLoss}) hold for any $p_i \in \R^+$. This property reflects the homogeneity seen by Baez et al. \cite{baez2011characterization}, and it appears to imply a usefulness beyond the theory of probability. We explore these ideas further in appendix \ref{SECTION_Simplex}.

We now show that $\mu$ can, in fact, be used to construct a signed measure space. In the next section we shall demonstrate that this measure space can be used to represent many information-theoretic quantities, including many which could not previously be accessed from the signed measure space perspective, and we show that it is indeed a refinement of the $I$-measure given by Yeung~\cite{yeung1991new}.

\begin{theorem}
\label{THEOREM_InteriorLossIsASignedMeasure}
Let $\Omega$ be a finite set of outcomes and let $\Sigma$ be the $\sigma$-algebra generated by all of the elements $b\in \Delta\Omega$. For $S\subseteq \Delta\Omega$ define $\mu(S) = \sum_{b\in S} \mu(b)$. Then $(\Delta\Omega, \Sigma, \mu)$ is a finite signed measure space.
\end{theorem}

\begin{proof}
Setting $\mu(\varnothing) = 0$, and using the definition of $\mu(S)$ we see that $\mu$ is at least countably additive across disjoint sets in $\Sigma$. Hence $(\Delta \Omega, \Sigma, \mu)$ is a signed measure space.
\end{proof}

Although we have shown that what we have constructed is, in fact, a signed measure space, we have not yet demonstrated that this space is consistent with the signed measure of Yeung, or that it can be used to represent any measure besides the entropy of a variable $H(X)$. Furthermore, we have not yet demonstrated that the M\"obius inversion is a reasonable approach for constructing a signed measure in this case. Indeed, given any system of objects, the M\"obius inversion could, in principle, be used to construct an additive function and, somewhat trivially, a signed measure on a corresponding space. That this function would have some intrinsic meaning is much harder to demonstrate. In this case, we now show that the measure $\mu$ has several analytic properties which seem to suggest a naturality to its construction. In the next section we also show that the measure $\mu$ has additional explanatory power (that is, it captures a larger class of information quantities).

We now briefly explore the properties of the total loss $L$ and the measure $\mu$. Some of these properties are quite intriguing; in particular the result of theorem \ref{AlternatingDerivatives} seems to imply a much more fundamental connection between the M\"obius inversion and Shannon entropy - so much so that its use seems quite justified.

\subsection{Properties of entropy loss, $L$}
\label{SUBSECTION_EntropyLossProperties}

The function $L$ has some properties that the entropy measure $H$ does not. It is true that for $\sum p_i = 1$ we have $L(p_1,\ldots, p_n) = H(p_1, \ldots, p_n)$, but this is not true if, as a function, we allow for the case when $\sum p_i \neq 1$.

The loss measure $L$ has some symmetry properties that $H$ lacks. In the classic paper of Shannon introducing his theory of communication \cite{shannon1948mathematical}, he introduces three requirements that the measure $H$ might naturally be expected to possess. The third of these is given as

\begin{quote}
\textit{If a choice be broken down into two successive choices, the original $H$ should be the weighted sum of the individual values of $H$.}
\end{quote}

As an example, Shannon gives
\begin{equation}
\label{EQUATION_Shannons3rd}
H\left(\frac{1}{2}, \frac{1}{3}, \frac{1}{6}\right) = H\left(\frac{1}{2}, \frac{1}{2}\right) + \frac{1}{2} H\left(\frac{2}{3}, \frac{1}{3}\right).
\end{equation}
What might bother us in this equation is the factor of $\frac{1}{2}$; it is an algebraic annoyance that in general
\begin{equation}
    k H(p_1, \ldots, p_n) \neq H(k p_1, \ldots, k p_n).
\end{equation}
In this scenario we are unable to remove this factor, and we are forced instead to keep track of multiple coefficients. Working with the entropy \textit{loss}, however, has a unique benefit:

\begin{proposition}
\label{PROPOSITION_LossFactors}
Let $p_1,\ldots, p_l \in \R^+$, and let $k \in \R^+$ where there is no constraint on $\sum p_i$. Then we have
\begin{equation}
k L(p_1, \ldots, p_l) = L( k p_1, \ldots, k p_l).
\end{equation}
That is, $L$ is homogeneous of order 1.
\end{proposition}
\begin{proof}
    \begin{align}
        L(kp_1,\ldots, kp_n) & = \sum_{i = 1}^n kp_i \log (kp_i) \notag \\
       & \quad - \left[ \sum_{i=1}^n kp_i \right] \log \left[k \sum_{i=1}^n p_i \right] \notag\\
        & = k \sum_{i=1}^n \left(p_i \log(p_i) + p_i \log (k)\right) \\
        & \quad - k \left[ \sum_{i=1}^n p_i \right]\left[ \log k + \log \left( \sum_{i=1}^n p_i \right) \right] \notag \\
        & = k L(p_1,\dots, p_n). \notag
    \end{align}
\end{proof}

This result can also be seen in the context of morphisms between probability measures the work on entropy loss by Baez et al. \cite{baez2011characterization}. Furthermore, Baez et al. also demonstrate the corresponding result for the Tsallis entropies \cite{tsallis1988possible, havrda1967quantification, patil1982diversity}:

\begin{theorem}
\label{THEOREM_LossFactorsTsallis}
Let $p_1,\ldots, p_l \in \R^+$, and let $k \in \R^+$ where there is no constraint on $\sum p_i$. Let $L_d$ be the $d$-th order Tsallis entropy loss. Then we have
\begin{equation}
k^d L_d(p_1, \ldots, p_l) = L_d( k p_1, \ldots, k p_l).
\end{equation}
That is, $L_d$ is homogeneous of order $d$.
\end{theorem}

\subsection{Properties of the measure $\mu$}

We now move on to the measure $\mu$ in the classical case (i.e. $d=1$). In this case, $\mu$ has some uniquely powerful analytic properties, some of which will be useful for proving other results, and others which may have applications to the study of bounding problems on information quantities. \revlbl{R1.1}\added{In particular, the measure of individual atoms is subject to constraints which are stronger than those on the $I$-measure.} We briefly state a result which gives a more explicit formula for the interior loss of a given atom.

\begin{lemma}[Interior loss identity]
\label{LEMMA_InteriorLossIdentity}
Let $T = \{p_1,\ldots, p_k\}$ be some collection of probabilities. For notational clarity we will write
\begin{equation}
\sigma(T) = \sigma(p_1,\ldots, p_k) = (p_1 + \cdots + p_k)^{(p_1 + \cdots + p_k)}.
\end{equation}
Further still we shall write
\begin{equation}
A_k = \prod_{\substack{S \subseteq\{p_1,\ldots, p_n\} \\ |S| = k}} \sigma(S).
\end{equation}
Then we have that
\begin{equation}
\label{ClearerExpression}
\mu(p_1,\ldots, p_n) = \sum_{k=1}^n (-1)^{n-k}\log (A_k)
\end{equation}
\end{lemma}

This lemma demonstrates that the atoms of our decomposition are measured by alternating sums of logarithms, justifying the name \textbf{logarithmic decomposition}. The next lemma allows for the confident inclusion of 0 in our domain for $\mu$.

\begin{lemma}[Interior loss at 0]
\label{InteriorLossAt0}
For $p_1,\ldots, p_n, x \in \R^+$ where $n\geq 0$, we have
\begin{equation}
\lim_{x\to 0} \mu(p_1,\ldots, p_n, x) = 0
\end{equation}
\end{lemma}

Because of this fact, we shall allow ourselves to extend the domain of $\mu$ to be defined for zero probabilities. This property is helpful, as in many cases it will allow us to ignore the contributions of various atoms where one of the associated probabilities is zero.

We will now proceed by showing the first of two peculiar and surprising properties of $\mu$.

\begin{lemma}
\label{LEMMA_MuAtInfinity}
Let $p_1,\ldots, p_{n-1}, x \in \R^+$ and let $x$ vary. Then
\begin{equation}
\lim_{x \to \infty} | \mu(p_1,\ldots, p_{n-1}, x)| = |\mu(p_1,\ldots, p_{n-1})|
\end{equation}
\end{lemma}

\begin{definition}
Let $b\in \Delta\Omega$. Then $b = \omega_1\omega_2 \ldots \omega_d$ for some $d\geq 1$. We define the \textbf{degree} of $b$ to be the number of outcomes it contains. That is, $\deg(b) = d$.
\end{definition}

This lemma reveals that the magnitude of a degree $d$ atom tends towards the magnitude of a degree $d-1$ atom when one of the arguments tends to infinity. While this could never happen in a probability space, the algebraic result holds nonetheless, and we will use it to construct the next few results, whose utility in usual probability spaces is much clearer. Geometrically speaking, this lemma says that the measure of a simplex will tend towards the measure of one of its edges when one of the ``probabilities'' grows towards infinity. 

The next theorem demonstrates the useful property that logarithmic atoms have an intrinsic sign, which is fixed depending only on the degree $d$.

\begin{theorem}
\label{AlternatingDerivatives}
Let $p_2,\ldots, p_{n} \in \R^+$ be a sequence of nonzero arguments for $n\geq 2$ and $m\geq 0$. Then
\begin{equation}
(-1)^{m+n} \frac{\partial^m \mu}{\partial x^m}(x, p_2,\ldots, p_n) \geq 0.
\end{equation}
\end{theorem}

Setting $m=0$ we immediately see that the sign of logarithmic atoms alternates solely on the number of outcomes they contain (its degree); a property which standard co-informations do not have. Stated otherwise: no knowledge of the underlying probabilities is needed to determine the sign of the measure of a given atom -- one only needs to know its degree.

Furthermore, the sign of these atoms \textit{and all of their derivatives in one argument} are completely fixed. This behaviour would not be expected if the choice to perform the M\"obius inversion were truly arbitrary. Rather, it shows that the entropy has the slightly surprising property that it behaves in a very specific way under this inversion.

This result also gives us monotonicity in each argument. Combining this with the bounding property of lemma \ref{LEMMA_MuAtInfinity}, we get the useful corollary:

\begin{corollary}[Interior magnitude can only decrease]
\label{COROLLARY_MuMagnitude}
Let $p_1,\ldots, p_{n-1}, \tau \in \R^+\cup\{0\}$ for $n\geq 3$. Then
\begin{equation}
|\mu(p_1,\ldots, p_{n-1}, \tau)| < |\mu(p_1,\ldots, p_{n-1})|
\end{equation}
\end{corollary}

This result is quite powerful in that it works for $p_1,\ldots, p_{n-1}, \tau \in [0,\infty)$. For our information-theoretical purposes, we will naturally require that $p_i\in [0,1]$, so the measure of successively higher-order atoms in $\Delta \Omega$ will in fact \textbf{strictly} decrease, with the slowest descent for $p_1 = \cdots = p_{n}$. Geometrically speaking, the contribution to the entropy of every simplex is bounded in magnitude by the contribution to the entropy of its boundaries, with equality for an infinite argument (which will not happen when locally studying random variables). The peculiarity that this is well-defined for all $p \in \R^+$ means that the logarithmic decomposition has a potentially useful application in the study of signed measures on simplices in general.

\subsection{Uniqueness of the Measure}

It is worth exploring that this signed measure space for entropy is unique in some key ways. We shall see that it forms the basis of a natural signed measure for the topology of a simplex where the measures of interiors are constructed explicitly from knowledge about weights at the vertices.

The next theorem is a re-statement of the main theorem of \cite{baez2011characterization} from the perspective of the interior loss. Given a measure $\mu$ which measures the interiors of simplices, under certain conditions it is possible to show that $\mu$ must be the interior loss given in this work.

\begin{proposition}
\label{PROPOSITION_UniquenessOfMeasure}
Let $\mu$ be a function assigning values to the interiors of a simplex $S$ as a function of weights \added{$p_i$} assigned to their corresponding vertices. Furthermore, require that
\begin{itemize}
    \item{$\mu$ is homogeneous of degree $d$\revlbl{R2.4, R2.5}\added{:}
    \added{
    \begin{equation}
    \lambda^d\mu(p_1,\ldots, p_n) = \mu(\lambda p_1,\ldots, \lambda p_n);
    \end{equation}
    }}
    \item{$\mu$ is additive across disjoint systems\footnote{\added{Baez et al. use $\oplus$ when referring to morphisms in their work \cite{baez2011characterization}. As we have swapped from entropy loss morphisms to sets, the disjoint union captures the idea appropriately.}}\revlbl{R2.4, R2.5}\added{:
    if $S_1, S_2 \subseteq S$ are subsets of the simplex $S$, categorically we have
    \begin{equation}
        \mu(S_1 \sqcup S_2) = \mu(S_1) + \mu(S_2);
    \end{equation}}}
    \item{$\mu$ is additive under composition (functoriality)\footnote{Baez et al. use `functoriality' in their original work \cite{baez2011characterization}. In that work, loss is additive over chains of data processing. Viewed in the reverse, the contribution to entropy should be additive under composition of distributions.}\revlbl{R2.4, R2.5}\added{:
    \begin{equation}
        \mu(p_1 + q_1, p_2, \ldots, p_n) = \mu(p_1, \ldots, p_n) + \mu(q_1,  \ldots, p_n) + \mu(p_1, q_1, p_2, \ldots, p_n);
    \end{equation}
    so that the measure is invariant when any outcome is split in the above fashion;}}
    \item{$\mu$ is continuous in its arguments \added{$p_1,\ldots, p_n$};}
\end{itemize}
Then $\mu$ is the interior loss of degree $d$ given in this work (up to a scaling factor), and the only function generating $\mu$ is $H_d$, the Tsallis entropy of order $d$.
\end{proposition}

This result, as stated, hinges mostly on the work of Baez et al. in that placing similar constraints on this measure of morphisms on the interior measure $\mu$ is sufficient to constrain $\mu$ to the specific form of interior entropy loss on a class of discrete measures on simplices. 

Our last result in this section shows that the measure-theoretic perspective is quite natural in that it implies two of these assumptions for free. As such, we are able to give a result about discrete measures on simplices in general.

\begin{theorem}
    Let $\mu_d$ be a signed measure on the interiors of a simplex which is homogeneous of degree $d$, assigning measures as a continuous function of weights assigned to corresponding vertices. Then $\mu_d$ is the interior loss of degree $d$ up to scaling factor $k$.
\end{theorem}
\begin{proof}
    It is sufficient to argue that a signed measure must be \textit{additive} and \textit{functorial} on its underlying space.

    A signed measure must by its very nature be additive on disjoint sets so that $\mu(S_1 \sqcup S_2) = \mu(S_1) + \mu(S_2)$. Furthermore, as a chain of sets
    \begin{equation}
    S_1 \supseteq S_2 \supseteq S_3 \supseteq \cdots \supseteq S_n
    \end{equation}
    gives the natural collection of disjoint sets
    \begin{equation}
    S_1 \setminus S_n = S_1\setminus S_2 \cup S_2 \setminus S_3 \cup \cdots \cup S_{n-1} \setminus S_n,
    \end{equation}
    a signed measure should also have the \textit{functoriality} property when framed as a `loss' between (something akin to) variables. Hence being a signed measure, homogeneous, and continuous in its arguments is sufficient to specify the measure in this work $\mu$.
\end{proof}

It is unclear what the consequences of this interpretation of entropy as the natural measure for a simplex might be. We hope that this simplified perspective of entropy as a somewhat natural `measure for measures' may provide some insight across multiple domains.

In the next section we shall demonstrate that the unique properties (the fixed parity nature of the atoms of the decomposition and the bounding of size) of the measure $\mu$ can be applied to the study of various information quantities which we call \textbf{logarithmically decomposable} quantities. That is, we show that the language we have constructed has much additional explanatory power above the prevailing measure of Yeung \cite{yeung1991new}.

\ifSubfilesClassLoaded{\bibliographystyle{plain} \bibliography{main} }{}
\end{document}

\section{Quantities of Information}
\label{SECTION_Quantities}

Having constructed the signed measure space $\Delta\Omega$, we shall now demonstrate its utility by characterising various variable-level information quantities, including the mutual information, co-information, G\'acs-K\"orner common information \cite{gacs1973common}, \revlbl{R2.1, R2.2}\added{minimum sufficient statistic} and the O-information of Rosas et. al \cite{rosas2019quantifying}. \added{In addition, we will use the logarithmic decomposition to explore the functional common information of James and Crutchfield \cite{james2017multivariate} and the minimally sufficient statistic (MSS) \cite{fisher1922mathematical, lehmann2011completeness}.} We shall see also that the logarithmic decomposition can account for an entire class of information quantities which we call \textbf{logarithmically decomposable} quantities, which we expect may contain many standard information quantities.

To start with, we will first explore mutual information and co-information; quantities which describe the prevailing $I$-measure of Yeung \cite{yeung1991new}. We will see that these two measures can be reinterpreted and represented by this logarithmic decomposition, and hence we shall show that the measure $\mu$ is a strict refinement of the $I$-measure. From there, we show that, in addition to these quantities, our decomposition can also describe the Gács-Körner \revlbl{R2.1}\added{common information} which \added{is} not derivable using the $I$-measure alone.

\subsection{Mutual, Conditional and Co-information}
\label{SUBSECTION_mutual_information}

Let $X$ and $Y$ be two variables defined on a common outcome space $\Omega$, where $X$ and $Y$ correspond to partitions of $\Omega$, where parts in the partition represent distinct events in each variable. If needed, we can take $\Omega$ to be the meet of the two partitions corresponding to $X$ and $Y$, i.e. the coarsest partition which is finer than the partitions of $X$ and $Y$, so that both may be described as partitions on $\Omega$.

The degree to which the two variables interact can be quantified in terms of their entropies via their mutual information, $I(X;Y)$, where
\begin{equation}
\label{EQN_MutualInformation}
I(X;Y) := H(X) + H(Y) - H(X,Y).
\end{equation}
The mutual information captures the degree to which knowledge of the variable $X$ reduces uncertainty about the variable $Y$, and vice versa. It is a strictly positive quantity, as $H(X,Y) \leq H(X) + H(Y)$, with equality when $X$ and $Y$ are independent. Several generalisations of the mutual information exist to more than two variables, but none have yet had the satisfactory ability to capture the notion of `information shared between three or more observers.' One possible generalisation of the mutual information for multiple variables is the \textbf{interaction information} or \textbf{co-information} \cite{mcgill1954multivariate, bell2003co}. This expression is defined recursively using the equation
\begin{equation}
I(X_1;\ldots ;X_r) = I(X_1;\ldots; X_{r-1}) - I(X_1;\ldots; X_{r-1} | X_r).
\end{equation}
The co-information is, algebraically, a very natural extension of the mutual information. An alternative derivation shows that the co-information is the result of applying the inclusion-exclusion principle to a system of variables $X_1,\ldots, X_r$ and combinations of joint entropies, so it is quite natural that it be represented as the central region of an $I$-diagram.

\begin{figure}[ht]
\begin{center}
\begin{tikzpicture}
    \newcommand{\growth}{0.8}
    
  % Circles
  \draw (0*\growth,0*\growth) circle (2);
  \draw (2*\growth,0*\growth) circle (2);
  \draw (1*\growth,1.732*\growth) circle (2);
  
  % Label in the middle
  \node at (1*\growth,0.5*\growth) {$I(X;Y;Z)$};
  \node at (-2*\growth-0.1, -2*\growth-0.1) {$X$};
  \node at (4*\growth+0.1, -2*\growth-0.1) {$Y$};
  \node at (1*\growth, 4*\growth + 0.6) {$Z$};
\end{tikzpicture}
\caption{The co-information between three variables.}
\end{center}
\end{figure}
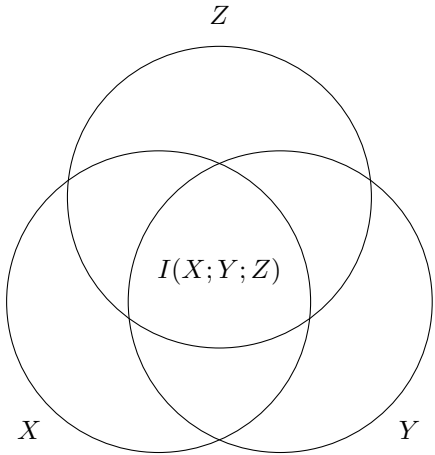

It would be perhaps reasonable to expect that the co-information should also be non-negative and represent shared information between three or more variables. Unfortunately, for three or more variables, the co-information $I(X_1;\ldots; X_n)$ can be both positive and negative, making it more difficult to interpret. A classic example of negative co-information is the XOR gate: $x, y, z \in \{0,1\}$ and $z = \mathrm{XOR}(x, y)$. In this system, equiprobable outcomes give $I(X;Y;Z)$ as $-1$ bits of information. In this case, the marginal mutual informations $I(X;Z)$ and $I(Y;Z)$ are zero, as knowledge of $X$ or $Y$ alone is not sufficient to deduce $Z$. Taken together, however, one is able to simply compute $Z$, so that $I(XY;Z) = 1$ bit. This effect, where deductive ability as a whole is greater than the sum of its parts, is known as \textbf{synergy}.

In general, the co-information is the sum of multiple kinds of information sharing effects. In systems of three variables, the co-information is precisely the sum of \textbf{synergistic} effects and \textbf{redundant} effects (where information can be thought of as being shared in a sense akin to the mutual information). The $I$-measure is unable to discern between these two effects. Other generalisations of the mutual information do exist, for example the \textbf{total correlation} \cite{watanabe1960information} and the \textbf{dual total correlation} \cite{te1978nonnegative}. However, both of these measures can be expressed as sums (possibly with multiplicity) of regions on $I$-diagrams, and hence also account for multiple sharing effects at once.

To start with, we would like to ensure our measure can at least represent the $I$-measure. From there, we will demonstrate the additional strengths of our decomposition's increased resolution. Our first definition will give us the connection between a random variable and a set representing its decomposition into atoms. Performing this construction will enable us to discuss co-information and regions in $I$-diagrams in terms of our decomposition atoms, while allowing us to explain how to explicitly represent abstract set-variables $\tilde{X}$.

\begin{definition}
\label{DEFINITION_content}
Given a random variable $X$, we define the \textbf{content} $\Delta X$ inside of $\Delta \Omega$ to be the set of all boundaries $b\in \Delta \Omega$ crossed by $X$. That is, if $X$ corresponds to a partition $P_1,\ldots, P_n$, then
\begin{multline}
\Delta X = \{ \text{$b_S: S \subseteq \Omega, \exists\,  \omega_i,\omega_j \in S$} \\
\text{with $\omega_i \in P_k$, $\omega_j\in P_l$ such that $k\neq l$ }\}.
\end{multline}
Intuitively, this means that at least two of the outcomes in $b_{\omega_1\ldots\omega_n}$ correspond to distinct events in $X$, although possibly more. We will in general make use of $\Delta$ to represent the logarithmic decomposition functor from random variables to their corresponding sets in $\Delta\Omega$. Under this correspondence, we have that the information quantity $H(X)$ is represented by the set $\Delta X$:
\begin{equation}
\label{EQUATION_EntropyContent}
    H(X) \longleftrightarrow \Delta X.
\end{equation}
We will see shortly that we need only measure $\Delta X$ to obtain $\mu(\Delta X) = H(X)$.
\end{definition}

\begin{remark}
It is straightforward to see how we can extend this to quantities like the mutual information. If mutual information reflects the inner region of an $I$-diagram between a pair of variables, then representing the content of two variables $X$ and $Y$ as $\Delta X$ and $\Delta Y$ should lead us quite naturally to the representation 
\begin{equation}
\label{EQUATION_MutualInformationContent}
I(X;Y) \longleftrightarrow \Delta X \cap \Delta Y.
\end{equation}
We make this construction more explicit in the proof of theorem \ref{THM_yeung_correspondence} below.
\end{remark}

We have now introduced the set $\Delta \Omega$, set representations $\Delta X$ for a given variable $X$ and we have explained how to measure the individual atoms in $\Delta \Omega$. However, we have not yet shown explicitly that
\begin{equation}
H(X) = \mu(\Delta X)
\end{equation}
or
\begin{equation}
I(X;Y) = \mu(\Delta X \cap \Delta Y).
\end{equation}
The theorem to follow will formalise this connection.

\begin{example}
To demonstrate our refinement, we consider the space $\Omega = \{1,2,3,4\}$. Let the partitions be given by $X =\{\{1, 3\}, \{2, 4\}\}$ and $Y = \{\{1, 2\}, \{3, 4\}\}$, as in figure \ref{FIG_two_random_vars}. In principle, we could also consider any other partitions of this outcome space $\Omega$. That is, our construction is only defined by the structure of the outcome space $\Omega$, not by the events defined upon it.

Taking the intersection of the contents $\Delta X \cap \Delta Y$ gives the content corresponding to $I(X;Y)$ as per equation \ref{EQUATION_MutualInformationContent}. These logarithmic atoms are given in in an $I$-diagram in figure \ref{XYLogarithmic}, with a representation of their corresponding entropic quantity given in figure \ref{XYLogarithmicPlus}.

\begin{figure}[ht]
\centering
    \scalebox{0.88}{
    \begin{tikzpicture}[scale=0.88]
    %%% XYZ with 1, 2, 3, 4
    \draw (0,0) -- (2, 0);
    \draw (0,0) -- (0, -2);
    \draw (0, -2) -- (2, -2);
    \draw (2, -2) -- (2, 0);
    \draw (0, -1) -- (2, -1);
    \draw (1, 0) -- (1, -2);
    \node at (0.5, -0.5) {$(1)$};
    \node at (1.5, -0.5) {$(2)$};
    \node at (0.5, -1.5) {$(3)$};
    \node at (1.5, -1.5) {$(4)$};
    \node at (1, -2.35) {$(X, Y)$};

    %%% XYZ in probabilities
    \draw (3,0) -- (5, 0);
    \draw (3,0) -- (3, -2);
    \draw (3, -2) -- (5, -2);
    \draw (5, -2) -- (5, 0);
    \draw (3, -1) -- (5, -1);
    \draw (4, 0) -- (4, -2);
    \node at (3.5, -0.5) {$0.1$};
    \node at (4.5, -0.5) {$0.2$};
    \node at (3.5, -1.5) {$0.3$};
    \node at (4.5, -1.5) {$0.4$};
    \node at (4, -2.35) {$(X, Y)$};

    %%% X
    \draw (6,0) -- (8, 0);
    \draw (6,0) -- (6, -2);
    \draw (6, -2) -- (8, -2);
    \draw (8, -2) -- (8, 0);
    \draw (7, 0) -- (7, -2);
    \node at (6.5, -1) {$0.4$};
    \node at (7.5, -1) {$0.6$};
    \node at (7, -2.35) {$X$};

    %%% Y
    \draw (9,0) -- (11, 0);
    \draw (9,0) -- (9, -2);
    \draw (9, -2) -- (11, -2);
    \draw (11, -2) -- (11, 0);
    \draw (9, -1) -- (11, -1);
    \node at (10, -0.5) {$0.3$};
    \node at (10, -1.5) {$0.7$};
    \node at (10, -2.35) {$Y$};

    \end{tikzpicture}
    }
\caption{Two random variables on the set $\Omega = \{1, 2, 3, 4\}$ with some illustrative probabilities.}
\label{FIG_two_random_vars}
\end{figure}
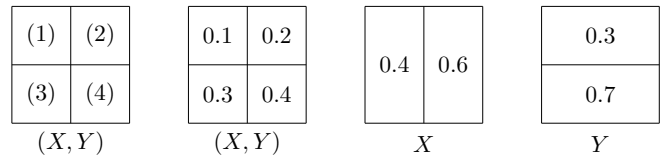

\begin{figure}[!t]
\centering
        \begin{tikzpicture}
        \newcommand{\spaceblubber}{0.35}
        \newcommand{\outerspace}{2.45}
        \newcommand{\innerspace}{0.9}
        \newcommand{\circlesize}{5cm}
        \newcommand{\circledisplacement}{0.75}
        \newcommand{\adjuss}{0.2}
        
        % Variable X
        \node [draw, circle, minimum size = \circlesize, label={135:$X$}] (X) at (-\circledisplacement, \adjuss){};
        
        % Variable Y
        \node [draw, circle, minimum size = \circlesize, label={45:$Y$}] (Y) at (\circledisplacement, \adjuss){};
        
        % Left components
        \node at (-\outerspace, \spaceblubber) {$b_{12}$};
        \node at (-\outerspace, -\spaceblubber) {$b_{34}$};
        % Right components
        \node at (\outerspace, \spaceblubber) {$b_{13}$};
        \node at (\outerspace, -\spaceblubber) {$b_{24}$};
        
        % Middle components
        \node at (0, 5*\spaceblubber) {$b_{14}$};
        \node at (0, 3*\spaceblubber) {$b_{23}$};
        \node at (-\innerspace, 1*\spaceblubber) {$b_{123}$};
        \node at (\innerspace, 1*\spaceblubber) {$b_{124}$};
        \node at (-\innerspace, -1*\spaceblubber) {$b_{134}$};
        \node at (\innerspace, -1*\spaceblubber) {$b_{234}$};
        \node at (0, -3*\spaceblubber) {$b_{1234}$};
        \end{tikzpicture}
        
        \caption{\textcolor{white}{Logarithmic atoms of $\Omega = \{1,2, 3, 4\}$ in the space $\Delta \Omega$.}}
        \label{XYLogarithmic}
\end{figure}
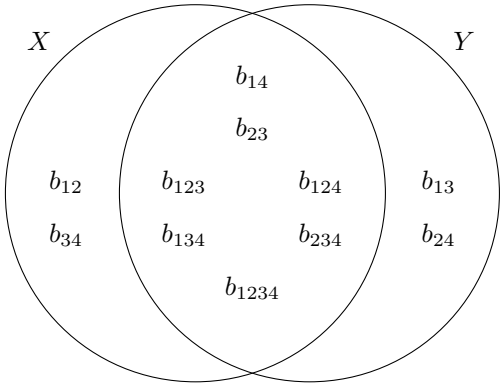

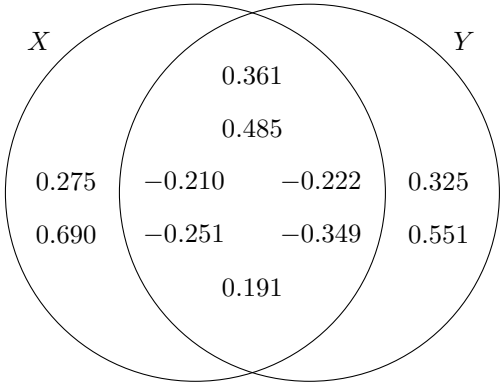
\begin{figure}[!t]
\centering
        \begin{tikzpicture}
        \newcommand{\spaceblubber}{0.35}
        \newcommand{\outerspace}{2.45}
        \newcommand{\innerspace}{0.9}
        \newcommand{\circlesize}{5cm}
        \newcommand{\circledisplacement}{0.75}
        \newcommand{\adjuss}{0.2}
        
        % Variable X
        \node [draw, circle, minimum size = \circlesize, label={135:$X$}] (X) at (-\circledisplacement, \adjuss){};
        
        % Variable Y
        \node [draw, circle, minimum size = \circlesize, label={45:$Y$}] (Y) at (\circledisplacement, \adjuss){};
        
        % Left components
        \node at (-\outerspace, \spaceblubber) {$0.275$};
        \node at (-\outerspace, -\spaceblubber) {$0.690$};
        % Right components
        \node at (\outerspace, \spaceblubber) {$0.325$};
        \node at (\outerspace, -\spaceblubber) {$0.551$};
        
        % Middle components
        \node at (0, 5*\spaceblubber) {$0.361$};
        \node at (0, 3*\spaceblubber) {$0.485$};
        \node at (-\innerspace, 1*\spaceblubber) {$-0.210$};
        \node at (\innerspace, 1*\spaceblubber) {$-0.222$};
        \node at (-\innerspace, -1*\spaceblubber) {$-0.251$};
        \node at (\innerspace, -1*\spaceblubber) {$-0.349$};
        \node at (0, -3*\spaceblubber) {$0.191$};
        \end{tikzpicture}
        
        \caption{The entropies associated to logarithmic atoms of $\Omega = \{1,2, 3, 4\}$ in the space $\Delta \Omega$. Note that, as per Theorem~\ref{AlternatingDerivatives}, the odd-degree atoms are negative.}
        \label{XYLogarithmicPlus}
\end{figure}

\end{example}

The next theorem is the main result of this paper, demonstrating that this logarithmic decomposition is consistent with the standard decomposition of Yeung \cite{yeung1991new}.

\begin{theorem}
\label{THM_yeung_correspondence}
Let $R$ be a region on an $I$-diagram of variables $X_1,\ldots, X_r$ with Yeung's $I$-measure. In particular, $R$ is given by some set-theoretic expression in terms of the set variables $\tilde{X}_1, \ldots, \tilde{X}_r$ under some combination of unions, intersections and set differences.

Making the formal substitution
\begin{equation}
    \tilde{X}_1, \tilde{X}_2, \ldots, \tilde{X}_r \quad \longleftrightarrow \quad \Delta X_1, \Delta X_2, \ldots, \Delta X_r \\
\end{equation}
to obtain an expression $\Delta R$ in terms of the $\Delta X_i$, we have
\begin{equation}
I(R) = \sum_{B \in \Delta R} \mu(B).
\end{equation}
That is, the interior loss measure $\mu$ is consistent with Yeung's $I$-measure.
\end{theorem}

\begin{remark}
In particular, we have the following identities:
\begin{equation}
H(X) = \mu(\Delta X)
\end{equation}
\begin{equation}
H(X,Y) = \mu(\Delta XY) = \mu(\Delta X \cup \Delta Y)
\end{equation}
\begin{equation}
I(X;Y) = \mu(\Delta X \cap \Delta Y)
\end{equation}
\end{remark}

We shall explore formal sums on the set $\Delta \Omega$ later, and we shall see that such a construction is able to characterise such quantities as the total correlation (TC) \cite{watanabe1960information} and the O-information \cite{rosas2019quantifying}, further expanding the range of quantities our decomposition can account for.

Previously we mentioned that the $I$-measure, while able to quantify the entropies of common information variables after they are found, does not provide any additional insight into their calculation. In the words of Gács and Körner in their paper introducing their common information, it appears to have `nothing to do with mutual information' as mutual information does not arise as the solution to a coding problem \cite{gacs1973common}. We now show that our decomposition is able to account for the Gács-Körner \added{common information}.

This provides not only an intuitive language for relating the mutual information to the common information, but also appears to have some explanatory power as to \textit{why} they do not appear to speak the same language.

\subsection{G\'acs-K\"orner Common Information}

An intrinsic problem in the study of random variables is that interactions between variables often (almost always) cannot be encoded with a third variable \cite{gacs1973common}. For instance, the G\'acs-K\"orner formulation of this \textbf{common information} has been shown to have little relation to the mutual information in most scenarios.

We have seen in section \ref{SUBSECTION_mutual_information} that mutual information, conditional entropies and the co-information can be neatly expressed as subsets of $\Delta\Omega$ and hence are captured by our decomposition. We will now demonstrate that the logarithmic decomposition is also able to describe the common information of G\'acs and K\"orner, which is a standard metric used to describe information that two variables jointly encode.

To do this, we shall demonstrate that this common information shared between a finite collection of variables $X_1,\ldots, X_r$ corresponds to a subset of $\Delta X_1 \cap \cdots \cap \Delta X_r$.

\begin{definition}[G\'acs-K\"orner Common Information]
The G\'acs-K\"orner common information on a finite set of random variables $X_1,\ldots, X_r$ \cite{gacs1973common} is given by
\begin{multline}
C_{\mathrm{GK}}(X_1;\ldots; X_r) = \max_{Z} H(Z) \\ \text{such that $f_1(X_1) = \cdots = f_r(X_r) = Z$ for some $f_i$.}
\end{multline}
\end{definition}

The common information quantifies interactions between variables which can be extracted and represented by another variable \cite{yu2016generalized}. That is to say, the G\'acs-K\"orner common information captures interactions between variables which are, in some sense, jointly \textit{encoding} certain events or outcomes as distinct from the others. The common information is upper-bounded by the mutual information between any pair of variables in $X_1,\ldots, X_n$, but is otherwise difficult to relate back to the mutual information in most cases.

\begin{theorem}
\label{THM_gacs_korner}
The G\'acs-K\"orner common information of a finite set of variables $X_i$ corresponds to the maximal subset $C$ of $\bigcap_{i} \Delta X_i$ such that there exists some random variable $Z$ with $\Delta Z = C$.
\end{theorem}

Intuitively speaking, we have that certain classes of subsets of $\Delta \Omega$ correspond to the entropy of variables, and some do not. That is to say, these is some class $\mathcal{S} \subseteq \mathcal{P}(\Delta \Omega)$ (where $\mathcal{P}$ is the power set) of sets which can be represented by variables, with remaining sets $\mathcal{S}^c$ not representable. We shall characterise the representable sets later in the algebraic discussion. For now, we make this notion concrete.

\begin{definition}
Given a subset $R \subseteq \Delta\Omega$, we say that $R$ is \textbf{representable}\footnote{In an earlier version of this work \cite{down2023logarithmic}, we called this property \textit{discernibility}.} if it corresponds to the content of any random variable $Z$ on the same outcome space $\Omega$.

Moreover, given any subset $S\subseteq \Delta\Omega$, let $\Rep(S) \subseteq S$ be the largest representable subset of $S$. We will call this the \textbf{maximal representable subset of $S$}.
\end{definition}

Note that $\Rep(S)$ is well defined as the trivial random variable is always representable in $S$, and we also have uniqueness of $\Rep(S)$ and the variable that it corresponds to, as $\Delta: X \to \mathcal{P}(\Delta\Omega)$ is injective on isomorphism classes of random variables. To see this, suppose that we take two non-isomorphic variables $Z_1$ and $Z_2$ which we assume to also have maximal contents $\Delta Z_1, \Delta Z_2 \subseteq S$, then $\Delta Z_1 Z_2$, the content of their joint distribution, would be a larger subset of $S$, contradicting their maximality.

\begin{remark}
As seen in theorem \ref{THM_gacs_korner}, $\mu[\Rep(\bigcap_i \Delta X_i)] = C_{\mathrm{GK}}(X_1;\ldots; X_r)$, the G\'acs-K\"orner common information.
\end{remark}

For an example illustrating this result geometrically, see figure \ref{FIGURE_triangles_intersect}.

\newcommand{\grayopac}{80}
\newcommand{\varlabdist}{0.2}
\newcommand{\chrr}{0.4}

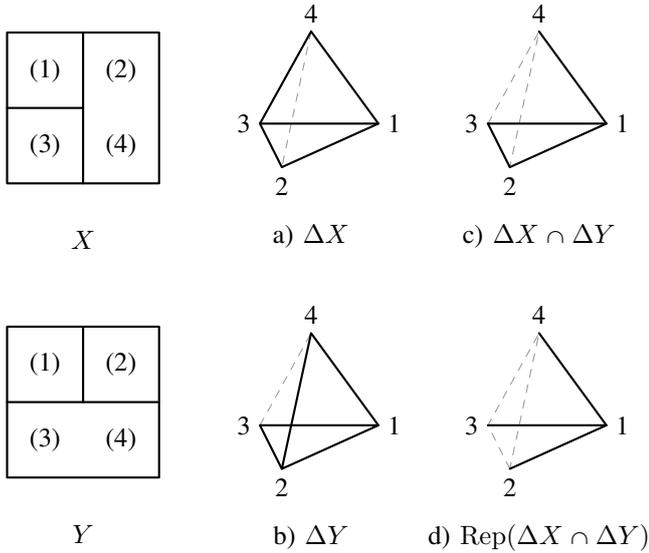
\begin{figure}[!t]
\centering
    \begin{tikzpicture}[line join = round, line cap = round]
    \begin{scope}[shift={(-10,1)}, scale=1]
    \scalebox{1.0}{
        %%% (X)
        \draw[thick] (6,0 + \chrr) -- (8,0 + \chrr) -- (8,-2 + \chrr) -- (6, -2 + \chrr) -- cycle;
        \draw[thick] (7,0 + \chrr) -- (7,-1 + \chrr) -- (6, -1 + \chrr);
        \draw[thick] (7, -1 + \chrr) -- (7, -2 + \chrr);
        \node at (6.5, -0.5 + \chrr) {(1)};
        \node at (7.5, -0.5 + \chrr) {(2)};
        \node at (6.5, -1.5 + \chrr) {(3)};
        \node at (7.5, -1.5 + \chrr) {(4)};
        \node at (7, -2.55 - \varlabdist + \chrr) {$X$};

        %%% (Y)
        \begin{scope}[shift={(0, -0.5)}]
        \draw[thick] (6,-3) -- (8,-3) -- (8,-5) -- (6, -5) -- cycle;
        \draw[thick] (7,-3) -- (7, -4);
        \draw[thick] (6, -4) -- (8, -4);
        \node at (6.5, -3.5) {(1)};
        \node at (7.5, -3.5) {(2)};
        \node at (6.5, -4.5) {(3)};
        \node at (7.5, -4.5) {(4)};
        \node at (7, -5.55 - \varlabdist) {$Y$};
        \end{scope}
        }
    \end{scope}
    \begin{scope}
        \coordinate [label=above:4] (4) at (0,{sqrt(2)},0);
        \coordinate [label=left:3] (3) at ({-.5*sqrt(3)},0,-.5);
        \coordinate [label=below:2] (2) at (0,0,1);
        \coordinate [label=right:1] (1) at ({.4*sqrt(3)},0,-.5);
    
        \begin{scope}
        \draw[dashed, gray!\grayopac] (4)--(2);
        \draw[thick, black] (1)--(3);
        \draw[thick, black] (2)--(1)--(4);
        \draw[thick, black] (3)--(2)--cycle;
        \draw[thick, black] (3)--(4)--cycle;
        \node at (0, -1.3) {a) $\Delta X$};
        \end{scope}
    \end{scope}
    \begin{scope}[shift={(0,-4)}]
        \coordinate [label=above:4] (4) at (0,{sqrt(2)},0);
        \coordinate [label=left:3] (3) at ({-.5*sqrt(3)},0,-.5);
        \coordinate [label=below:2] (2) at (0,0,1);
        \coordinate [label=right:1] (1) at ({.4*sqrt(3)},0,-.5);
    
        \begin{scope}
        \draw[dashed, gray!\grayopac] (3)--(4);
        \draw[thick, black] (1)--(3);
        \draw[thick, black] (2)--(1)--(4)--cycle;
        \draw[thick, black] (3)--(2)--cycle;
        \node at (0, -1.3) {b) $\Delta Y$};
        \end{scope}
    \end{scope}
    \begin{scope}[shift={(3,0)}]
    
        \coordinate [label=above:4] (4) at (0,{sqrt(2)},0);
        \coordinate [label=left:3] (3) at ({-.5*sqrt(3)},0,-.5);
        \coordinate [label=below:2] (2) at (0,0,1);
        \coordinate [label=right:1] (1) at ({.4*sqrt(3)},0,-.5);
    
        \begin{scope}
        \draw[dashed, gray!\grayopac] (3)--(4)--(2);
        \draw[thick, black] (1)--(3);
        \draw[thick, black] (2)--(1)--(4);
        \draw[thick, black] (3)--(2)--cycle;
        \node at (0, -1.3) {c) $\Delta X \cap \Delta Y$};
        \end{scope}
    \end{scope}
    \begin{scope}[shift={(3, -4)}]
        \coordinate [label=above:4] (4) at (0,{sqrt(2)},0);
        \coordinate [label=left:3] (3) at ({-.5*sqrt(3)},0,-.5);
        \coordinate [label=below:2] (2) at (0,0,1);
        \coordinate [label=right:1] (1) at ({.4*sqrt(3)},0,-.5);
    
        \begin{scope}
        \draw[dashed, gray!\grayopac] (2)--(3)--(4)--cycle;
        \draw[thick, black] (2)--(1)--(4);
        \draw[thick, black] (1)--(3);
        \node at (0, -1.3) {d) $\Rep(\Delta X\cap \Delta Y)$};
        \end{scope}
    \end{scope}
\end{tikzpicture}
\caption{The 1-dimensional atomic contents of $X$ and $Y$ and their (c) intersection and (d) maximal representable subset. Higher dimensional interactions are not displayed for clarity, but these faces and volumes intersect similarly.}
\label{FIGURE_triangles_intersect}
\end{figure}

We now move on to \revlbl{R2.1}\added{consider the functional common information \cite{james2017multivariate} and minimally sufficient statistic \cite{fisher1922mathematical,lehmann2011completeness}} and the O-information of Rosas et al. \cite{rosas2019quantifying}.

\revlbl{R2.1}
\subsection{\added{Functional common information}}

\added{The functional common information, defined in a footnote in the work where James and Crutchfield which introduced the Dyadic and Triadic systems \cite{james2017multivariate}, captures the minimum amount of entropy required to render a system of variables conditionally independent.}
\begin{definition}
    \added{Given a set of random variables $X_1,\ldots, X_n$, the \textbf{functional common information} \cite{james2017multivariate} is given by}
    \begin{equation}
    F[\{X_i\}] = \min_{\substack{\indep X_i | V \\ V = f(\{X_i\})}} H(V).
    \end{equation}
    \added{That is, the smallest entropy of a variable $V$, a function of the $X_i$, which renders the $X_i$ conditionally independent.}
\end{definition}

\begin{proposition}
\label{PROPOSITION_functional_common_info}
    \added{The functional common information is logarithmically decomposable without refinement, but not lattice decomposable.}
\end{proposition}

\added{To show how this can be calculated in practice, we give an example.}
\begin{example}
\label{EXAMPLE_FunctionalCommonInformation}
Let $X$ and $Y$ be defined on a common outcome space $\Omega$, where $\Omega = \{1, 2, 3, 4\}$. Let $X$ be given by the partition $\{\{1\}, \{2, 3, 4\}\}$ and $Y$ by the partition $\{\{2\}, \{1, 3, 4\}\}$. Then
\begin{equation}
\Delta X = \{12, 13, 14, 123, 124, 134, 1234\}
\end{equation}
and
\begin{equation}
\Delta Y = \{12, 23, 24, 123, 124, 234, 1234\}
\end{equation}
giving us
\begin{equation}
\Delta X \cap \Delta Y = \{ 12 , 123, 124, 1234 \}
\end{equation}
as a set. In this case, the problem of computing the \added{functional common information} is to find the lowest-entropy set $I \supseteq \Delta X \cap \Delta Y$ where $I$ is representable. \added{The minimising variable $V$ will then have $\Delta V = I$.}

Intuitively, given a boundary $12$, we know that in order to extend to a representable set, we must extend the boundary $12$ so that can be expressed in the form $\Delta V$ for some variable $V$. That is to say, we must find the lowest-entropy partition with $1$ and $2$ in separate parts. In this case we may assume that $3$ and $4$ are contained in the same part\added{, as separating them will certainly increase the entropy of the representative variable $V$}. 

The valid partitions in this case are themselves $X: \{\{1\}\{2, 3, 4\}\}$ and $Y: \{\{2\}, \{1, 3, 4\}\}$. The partition $\{\{1\},\{2\}, \{3,4\}\}$ is strictly more informative then both of these, so we need not consider it. Some atoms are common to both partitions, so will not be relevant for us to select the partition of greatest entropy. The atoms we can comfortably ignore are $12, 123, 124$, and $1234$.

Hence it suffices to select the smaller of $\mu( \{13, 14, 134\})$ or $\mu(\{23, 24, 234\})$. In particular, if $p(\omega_1) < p(\omega_2)$ then we select the former\added{, and if $p(\omega_2)< p(\omega_1)$ then we select the latter}.
\end{example}

\subsection{\added{The minimum sufficient statistic}}

\revlbl{R2.1, R2.2}\added{Fisher introduced the notion of a sufficient statistic back in 1922 \cite{fisher1922mathematical}. Namely, when a statistic is computed from observation in order to estimate a parameter, the statistic provides the maximum amount of information which can be garnered about the parameter from the data. More precisely:}

\begin{definition}
    \added{A statistic $T(X_1,\ldots, X_n)$ is \textbf{sufficient} for a parameter $\theta$ if the conditional distribution of $\bm{X} = (X_1,\ldots, X_n)$ given $T$ does not depend on $\theta$ \cite{fisher1922mathematical}.}
\end{definition}

\added{This notion, whereby $T(X)$ captures all available deductive information about $\theta$ from $\bm{X} = X_1,\ldots, X_n$, can be more carefully refined into the \textbf{minimum sufficient statistic} \cite{lehmann2011completeness}.}

\begin{definition}
    \added{A sufficient statistic $T(\bm{X})$ for $\theta$ is \textbf{minimal} if, for every sufficient statistic $G$, we have $T = f(G)$ for some function $f$.}
\end{definition}

\begin{proposition}
\label{PROPOSITION_minimum_sufficient_stat}
    \added{The minimally sufficient statistic is logarithmically decomposable without refinement, but not lattice decomposable.}
\end{proposition}

\subsection{Quantities with information multiplicity}

Thus far we have only explored information quantities which do not count any atoms with multiplicity. Many useful information quantities do not have this property. Two natural examples are the total correlation (TC), and the O-information of Rosas et al. \cite{watanabe1960information, rosas2019quantifying}. The O-information is useful for quantifying synergy and redundancy effects in multivariate systems, where it is used to determine if information representations are \textit{redundancy} or \textit{synergy dominated} \cite{rosas2019quantifying}. It has found much use in the study of information dynamics in brain networks \cite{stramaglia2021quantifying}, and has applications to detecting significant interactions between variables \cite{marinazzo2022information}.

If we are taken to counting atoms with multiplicity, we can extend the logarithmic decomposition so that it is able to capture these metrics. In particular, if we consider the natural extension
\begin{equation}
\Z \Delta \Omega = \left\{ \sum_{b \in \Delta \Omega} n_b b: n_b\in \Z \right\},
\end{equation}
expressions in $\Z\Delta\Omega$ will now correspond to expressions of entropy, counting atoms multiple times. Note that in this case, due to the additivity of the measure, $\mu(n_b b) = n_b \mu(b)$ as one would expect.

\begin{definition}
Let $\Omega$ be a discrete space of outcomes\footnote{It appears reasonable in this case to allow discrete variables where $\Omega$ is countable. A natural extension to all spaces $\Omega$ might be reasonable, but makes the intuition of logarithmic decomposition somewhat difficult.}. Let $X_\alpha: \alpha \in A$ be the family of variables corresponding to all possible partitions of $\Omega$. Then we call any finite sum
\begin{equation}
\sum_{\alpha \in A} n_\alpha H(X_\alpha)
\end{equation}
with $n_\alpha \in \Z$ an \textbf{entropy expression on $\Omega$}. Note that only finitely many terms have nonzero coefficient.
\end{definition}

\begin{proposition}
\label{PROPOSITION_EntropyExpressions}
There is a one-to-one correspondence
\begin{center}
\begin{tabular}{c}
$\left\{ \text{Unique entropy expressions with multiplicity on $\Omega$} \right\} $\\
$\updownarrow $\\
$\Z\Delta\Omega$\\
\end{tabular}
\end{center}
where we say two entropy expressions are the same if their value is identical on all underlying probability distributions on $\Omega$.
\end{proposition}

\begin{example}
Consider $\Omega = \{a, b, c, d\}$ with the partition corresponding to a variable $X$ given by $\{\{a,b\}, \{c, d\}\}$ and the partition for a variable $Y$ given as $\{\{a, c\}, \{b, d\}\}$. Then the entropy expression
\begin{equation}
I(X;Y) - H(X|Y) + H(X,Y)
\end{equation}
corresponds to the element
\begin{equation}
Z = ab + cd + 2ad + 2bc + 2abc + 2abd + 2acd + 2bcd + 2abcd
\end{equation}
inside of $\Z\Delta\Omega$. Measuring the expression, $\mu(Z)$, will give the entropy expression above for $X$ and $Y$, regardless of the underlying probability distribution.
\end{example}

We give the following brief expressions:

\begin{proposition}
Each of the following information quantities has a representation as follows.
\begin{enumerate}
\item{\textbf{Dual total correlation (DTC)}:
\begin{equation}
\DTC(\bm{X}_n) = \mu\left[\bigcup_{i,j} (\Delta X_i \cap \Delta X_j) \right]
\end{equation}}
\item{\textbf{Total correlation (TC)}:
\begin{equation}
\TC(\bm{X}_n) = \mu\left[\left(\sum_{i = 1}^n \Delta X_i\right) - \bigcup_{i=1}^n \Delta X_i \right]
\end{equation}}
\item{\textbf{The O-information}\footnote{Note that here we use $\Omega$ in the sense described by Rosas et al. in \cite{rosas2019quantifying}. It does not represent the outcome space as we have been using thus far.}:
\begin{equation}
\Omega(\bm{X}_n) = \DTC(\bm{X}_n) - \TC(\bm{X}_n).
\end{equation}
}
\end{enumerate}
\end{proposition}
\begin{proof}
It can be confirmed via symbolic substitution that the first two measures agree with the classical definition. The O-information is defined as the difference between the dual total correlation and the total correlation, so this suffices.
\end{proof}

\subsection{Logarithmically decomposable quantities}

Throughout this work we have seen multiple quantities which can be expressed using the logarithmic decomposition over an outcome space $\Omega$. There seems to be a subtle distinction between these representations, however, and we should treat it carefully. Most quantities examined in this work, namely, entropy, mutual information, co-information, and the Gács-Körner common information, meanwhile, can be derived using only the \textbf{lattice}-theoretic data (arguably even the set-theoretic data is sufficient). The \added{functional} common information was a little different; it required both the lattice-theoretic data \textit{and knowledge of the underlying probabilities}. That is, the \added{functional common information} is not a purely lattice-based measure. We will now briefly define two new ideas: \textbf{logarithmic decomposability} and \textbf{lattice decomposability}.

\begin{remark}
We will use the notation $\nabla^n$ to represent the simplex of probabilities in the space $\Omega$ with $|\Omega| = n$. We do this to avoid conflict with our use of $\Delta$.
\end{remark}

We first give a definition to represent any quantity which has a representation as a set in $\Delta \Omega$, possibly dependent on the underlying probabilities (\added{as we saw with the functional common information from \cite{james2017multivariate}}).

\begin{definition}
Given a collection of random variables $\{X_\alpha: \alpha \in A\}$ for some index set $A$ on a common outcome space $\Omega$, we let $\mathcal{A} = \mathcal{P}(A)$ be the powerset of $A$, and we define a \textbf{variable quantity} to be any map $f: \nabla^n \times \mathcal{A} \to \R$, so that $f$ might also explicitly depend on the underlying probability distribution.
\end{definition}

Now we give a definition which captures the idea that, for large, continuous areas in the simplex $\nabla^n$, there is a stable representation of the variable quantity as some subset $\Delta \Omega$. This might change as we alter the variables, but mostly it is stable with small changes to the input.

\begin{definition}
Let $f: \nabla^n \times \mathcal{A} \to \R$ on $\Omega$ be a variable quantity with $|\Omega| = n$. Suppose there exists a piecewise continuous function
\begin{equation}
f^*: \nabla^n \times \mathcal{A} \to \Delta \Omega
\end{equation}
where $\nabla_n$ is the probability simplex on $n$ outcomes and we equip $\mathcal{A}$ and $\Delta \Omega$ with the discrete topology and $\nabla^n$ with the usual Euclidean topology, where
\begin{equation}
f((p_1,\ldots, p_n), A) = \mu[f^*((p_1,\ldots, p_n), A)].
\end{equation}
Then we will say that $f$ is \textbf{logarithmically decomposable}.
\end{definition}

In this sense, we have accounted for the fact that the \added{functional common information} has two stable representations in $\Delta \Omega$ depending on the underlying probabilities. It is no surprise that all quantities given thus far have this property. Perhaps more interesting are those properties which are logarithmically decomposable but \textit{always} have a stable representation in $\Delta \Omega$.

\subsection{Lattice-decomposable quantities}

We have seen that the \added{functional} common information can change representation in $\Delta\Omega$ based on the underlying probability distribution. In terms of performing set-theoretic information theory, this seems to imply that the construction of the \added{functional} common information requires knowledge over and above variables $\Delta X_i$ as sets -- that is, it does not seem particularly natural from the set-theoretic perspective. We give the natural extension.

\begin{definition}
Let $f: \nabla^n \times \mathcal{A} \to \R$ on $\Omega$ be a variable quantity with $|\Omega| = n$. Suppose there exists a function
\begin{equation}
f^*: \mathcal{A} \to \Delta \Omega
\end{equation}
where
\begin{equation}
f((p_1,\ldots, p_n), A) = \mu[f^*(A)].
\end{equation}
Then we will say that $f$ is \textbf{lattice-decomposable}.
\end{definition}

This captures the idea that the function $f$ can be evaluated by first computing the logarithmic decomposition (which remains stable for all underlying probability distributions) and then applying the measure $\mu$. This is a stronger property, and certainly any lattice-decomposable quantity is logarithmically decomposable.

We have seen that all entropies, mutual informations and co-informations are lattice-decomposable, as they require no reference to the underlying probability distribution. Moreover, we also saw that the G\'acs-K\"orner common information has this property, where the \added{functional} common information does not.

For any quantities where we have logarithmic decomposability, we are able to understand now better the relationship between these quantities. Now, for example, we can explain the fact that common information is \textit{much less} than mutual information (\cite{gacs1973common}) in terms of mutual information not being representable -- that it lacks certain atomic supports which are necessary for representation with a variable. One could easily formulate a counting argument to quantify how many entropy expressions are representable and how many are not. For any logarithmically decomposable quantity, we can explain the negativity of certain information quantities in terms of the signs of their atoms, and use this to infer something about the qualities behind their representations. Many variable interactions can now be seen through a common lens, where we can break them down into their constituent \textit{atoms}.

In this section we have seen that logarithmically decomposable quantities appear to speak the same language; even if that language was previously unseen when using the coarse perspective of the $I$-measure. Learning to speak this new language, like any language, might bring us many new perspectives on old concepts.

\ifSubfilesClassLoaded{\bibliographystyle{plain} \bibliography{main} }{}
\end{document}

\section{Behaviour under Refinements}
\label{SECTION_Inclusions}

\subsection{Refinements of $\Omega$ without refining variable partitions}

All of the exploration thus far has only dealt with discrete probability spaces, and, moreover, only on spaces where the (joint) outcome space is specified. In order to construct a meaningful extension of the measure $\mu$ to continuous probability spaces, we will need to understand how the measure interacts with partition refinement, and hence explore how this might behave as we take successively finer and finer outcome spaces. Having explored this, we will then construct the direct limit of these objects to extract a continuous construction of $\Delta X$, which we have labelled $\delta X$.

Although it would be computationally challenging to compute the measures of all atoms for fine-grained systems, we demonstrate that the constructed space does, at least algebraically, deal with the interaction of arbitrarily many variables in a fashion which still has the structure of a measurable space. We leave explorations of the structure of this space to future work.

\begin{definition}
\label{phi-psi}
Let $\Omega$ be a set of discrete outcomes and let $\Omega'$ be a refinement of $\Omega$, such that for each $\omega \in \Omega$ there is some corresponding finite set $\{\omega'_1, \ldots \omega'_k\}$ partitioning $\omega$. For each such partition we shall write $\phi(\omega) = \{\omega'_1,\ldots, \omega'_k\}$. Hence we could consider
\begin{equation}
\phi: \Omega \to \Omega'
\end{equation}
as a mapping between sets (in practice, this could be viewed as a non-injective function $\phi\inv: \Omega' \to \Omega$). We refer to these mappings as \textbf{refinements}.
\end{definition}

We would like to be certain that in the case that the partition of a variable remains unchanged, that its representation in the refinement $\phi$ does not interfere with its measure. For example, given a variable $X$ defined with partition $\{\{a, b\}, \{c\}\}$ on $\Omega$, and a refinement splitting $a$ into $a_1$ and $a_2$,
\begin{equation}
Q_X = \{\{a, b\}, \{c\}\} \mapsto \{\{a_1, a_2, b\}, \{c\}\} = \phi(Q_{X}) = Q_{X'},
\end{equation}
we expect that $\mu(X)$ should be equal to $\mu(X')$. This is in fact the case.

\begin{proposition}
\label{PROPOSITION_EventEntropyStableUnderRefinement}
Let $Q_X$ be the partition of a variable $X$ on an outcome space $\Omega$, and let $\phi(Q_{X})$ be the image of this partition under a refinement $\phi$ as above. Abusing notation, we have that
\begin{equation}
\mu(\Delta Q_X) = \mu(\Delta \phi(Q_X)).
\end{equation}
That is, the measure $\mu$ is invariant under refinements up to partition.
\end{proposition}
\begin{proof}
As $\mu(\Delta Q_X) = \mu(\Delta X) = H(X)$, and the probabilities of given events in $X$ is invariant up to partition under $\phi$, we know that $\mu(\Delta Q_X) = H(X) = \mu(\Delta \phi(Q_X))$.
\end{proof}

In this case we are refining the space $\Omega$, but \textit{not} refining the partition corresponding to $X$, which is contrary to what might be expected if we were to perform a limiting process to extract a continuous variable. Intuitively speaking, the purpose of this result is to show that, provided that the partition studied is unchanged, the measure will not be affected by the symbols we use to describe it; it remains stable before and after the refinement.

\added{
\begin{definition}
    We use the notation $\Z\Delta\Omega$ to mean the set of all formal sums
    \begin{equation}
    \sum_{b \,\in\, \Delta\Omega} k_b b
    \end{equation}
    where the $k_b$ are taken from $\Z$. It is convenient to extend the measure to elements in this set by allowing
    \begin{equation}
    \mu\left(\, \sum_{b\,\in\,\Delta\Omega} k_b b \,\right) = \sum_{b\in \Delta\Omega} k_b \cdot \mu(b),
    \end{equation}
    as is most natural.
\end{definition}}

For completeness, we now give a result that shows the truly scaleless nature of our decomposition.

\begin{definition}
\label{DEFINITION_ThreeOperators}
For the purposes of the following result, we give enhanced definitions for three operators: contents $\Delta \cdot$, refinements $\phi(\cdot)$ and restrictions $\cdot|_S$.
\begin{itemize}
\item{Let $P$ be a partition of some set $\Omega$ (not necessarily taken to represent an outcome space). As before, we let $\Delta P$ be the set of all subsets \replaced{$W \subseteq \Omega$}{$S \subseteq \Omega$} which cross a boundary in $P$.}
\item{Let $\phi$ be a finite refinement from $\Omega$ to $\Omega'$ (i.e. $|\Omega'|$ is finite). We will let $\phi$ act on a partition $P$ by re-expressing it in $\Omega'$, so that $\phi(P) = P'$. If $\omega \mapsto \{\omega_a, \omega_b\}$ in the refinement $\phi$, we let we let $\phi$ act on elements of $\Z\Delta\Omega$ by sending each atom $\omega_1 \ldots \omega_n \omega$ to $\omega_1 \ldots \omega_n \omega_a + \omega_1 \ldots \omega_n \omega_b + \omega_1\ldots \omega_n \omega_a \omega_b$, possibly expanding multiple times. For more complex expressions, we let $\phi$ act additively across elements of $\Z\Delta\Omega$.}
\item{Let $S$ be some subset of the set $\Omega$. We write $P|_S$ to mean the partition $P$ restricted to the subset $S$. In particular, $P|_S$ is a partition of $S$. Given some element $Z \in \Z\Delta\Omega$, we send it to its image $Z|_S$ by removing all atoms containing outcomes not contained in $S$.}
\end{itemize}
\end{definition}

\begin{theorem}
\label{THEOREM_ThreeOperators}
Let $P$ be a partition of a set $\Omega$ (not necessarily taken in this context to represent the entire outcome space). Let $\phi$ be a refinement into finitely many parts, and let $S$ be a subset of $\Omega$ to which we will restrict. Then the three operations $\Delta \cdot$, $\phi(\cdot)$ and $\cdot |_S$ all commute.

\revlbl{R2.7}\added{In particular, for any partition $P$ defined on $\Omega$ and a subset $S$ defined on $\Omega$, we have:}
\begin{align}
\phi(\Delta P) &= \Delta \phi(P)\,; \\
\Delta P|_S &= \Delta(P|_S)\,; \text{\quad and} \\
\phi(P)|_S &= \phi(P|_S).
\end{align}
\added{In addition, for any content $C \in \Z\Delta\Omega$, we have}
\begin{equation}
    \phi(C)|_S = \phi(C|_S).
\end{equation}
\end{theorem}

This theorem shows that much of our thinking is indeed consistent; the content operator and refinement operator play well with restriction, as we would expect them to with the scalelessness of our decomposition. Given this additional power, it is only right to strengthen proposition \ref{PROPOSITION_EventEntropyStableUnderRefinement} to all subsets of $\Delta \Omega$.

\begin{corollary}
\label{COROLLARY_AllRefinementsStable}
Let \replaced{$C$}{$S$} be an element of $\Z\Delta\Omega$ for some finite outcome space $\Omega$ and let $\phi$ be a refinement of $\Omega$. Then
\begin{equation}
\mu(C) = \mu(\phi(C)).
\end{equation}
\end{corollary}
\begin{proof}
By proposition \ref{PROPOSITION_EntropyExpressions}, we have that every element of $\Z\Delta\Omega$ corresponds to a unique entropy expression. As all entropy expressions are stable under refinement (their partitions are stable), this follows immediately from proposition \ref{PROPOSITION_EventEntropyStableUnderRefinement}.
\end{proof}

\subsection{Refinements of $\Omega$ with refinement of variable partitions}

The arguably  more interesting case is when refining the outcome space $\Omega$ will allow us to gain an increased resolution in $X$, as is the case for when one wishes to study continuous variables in general by approximating them with discrete variables. In this case, the continuous variable $X$ is discretised into `bins' with a discrete probability, but making these bins smaller (refining $\Omega$) will then correspond to refining the partition of $X$ also.

In this case we expect that the measure of $\Delta X$ will increase under refinements, as would normally be expected when introducing a finer granularity. This corresponds loosely to the limiting process of Jaynes \cite{jaynes1957information, jaynes1968prior}, where classically refining will lead to an additional $\log N$ term in the calculation. 

In order to discuss continuous variables, we will construct an equivalence on sets $S \subseteq \Delta \Omega$ following some refinement $\Omega \to \Omega'$, where the space $\Omega$ is refined but the underlying partition $S$ is not. Using this we shall construct the direct limit, and use this in the next section to explore descriptions for continuous variables. Constructing this relation will allow us to logarithmically decompose while being more agnostic about the choice of granularity -- provided a sufficiently fine outcome space $\Omega$ is chosen, we can represent all possible partitions of interest.

\begin{definition}
Let $\mathcal{T}$ be the set of all possible finite partitions of $\Omega$. Note, in particular, that we allow these to be arbitrarily fine.

Let $T_1$ and $T_2$ be two finite partitions in $\mathcal{T}$. If $T_2$ is a refinement of $T_1$, then there exists a mapping $\psi_{T_1 \to T_2}$ sending sets in $\Delta T_1$ to sets in $\Delta T_2$ (as discussed in the previous section). Recall also that the partition corresponding to the joint discrete variable $T_1 T_2$ will be finer than both $T_1$ and $T_2$.

Then, given two subsets $S_1 \subseteq \Delta T_1$ and $S_2 \subseteq \Delta T_2$, we say that $S_1$ is \textbf{equivalent} to $S_2$ and write $S_1 \sim S_2$ if there exists some partition $T$ finer than $T_1T_2$ with $\psi_{T_1 \to T} (S_1) = \psi_{T_2 \to T} (S_2)$. That is, the image of $S_1$ and $S_2$ is equal under a sufficient refinement.
\end{definition}

\begin{proposition}
The relation $\sim$ is an equivalence relation.
\end{proposition}
\begin{proof}
It is immediately clear that the relation is symmetric and reflexive. To see transitivity, consider $S_1 \sim S_2$ and $S_2 \sim S_3$. Then 
\begin{align}
\psi_{T_1 \to T_1T_2}(S_1) &= \psi_{T_2 \to T_1 T_2}(S_2), \quad\text{and} \notag \\
\psi_{T_2 \to T_2T_3}(S_2) &= \psi_{T_3 \to T_2T_3}(S_3).
\end{align}
As we have equality we may further state
\begin{align}
    \psi_{T_1 \to T_1T_2T_3} (S_1) &= \psi_{T_2 \to T_1T_2T_3} (S_2), \quad \text{and} \notag \\
    \psi_{T_2 \to T_1T_2T_3} (S_2) &= \psi_{T_3 \to T_1T_2T_3} (S_3).
\end{align}
That is, the images of $S_1, S_2$ and $S_3$ are all equal in $T_1 T_2 T_3$, giving us the equivalence of $S_1$ and $S_3$. 
\end{proof}

With this notion of equivalence under refinement, we will now construct a direct limit, with which we can begin to discuss continuous variables.

\begin{definition}
Let $\delta \Omega$ be the set
\begin{equation}
\{\bm{S} : \text{ $\bm{S}$ is an equivalence class under $\sim$}\}
\end{equation}
Where it is always possible to compare two complexes $\Delta X$ and $\Delta Y$ by considering their mappings into $\Delta XY$.

Note that the construction of $\delta \Omega$ is now being done in terms of sets rather than atoms. If we were to use atoms, then these atoms would then be represented by sets upon refinement, and it is sufficient to represent an atom with a singleton set.
\end{definition}

This construction can also be viewed as the direct limit on the directed set of partition refinements. Given a `global' outcome space $\Omega$, there are finite partitions $\Omega_i, i\in I$ with the property that the morphisms compose appropriately, i.e. $\psi_{\Omega_j \to \Omega_k} \circ \psi_{\Omega_i \to \Omega_j} = \psi_{\Omega_i \to \Omega_k}$.

We note that with this notation we can arbitrarily refine the expressions we've already obtained to finer and finer partitions. This construction is analogous to the construction of the rational numbers, and hence full treatment requires a completion step.

\begin{example}
Consider the two partitions in figure \ref{FIGURE_Inclusions} corresponding to two variables $X$ and $Y$, where $Y$ is strictly finer than $X$.

Here we start with an outcome space $\Omega$, sufficient to describe $\Delta X$ but not $\Delta Y$. As we refine $\Omega$ to $\phi(\Omega)$, we acquire enough resolution to describe $Y$, and we are still able to describe $\Delta X$. Under the refinement, $\Delta X$ and $\psi(\Delta X)$ are equivalent.

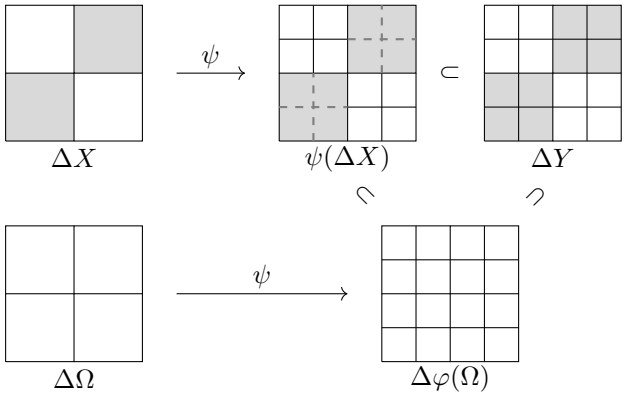
\begin{figure}[h]
\centering
\begin{tikzpicture}[scale=0.9]
    \begin{scope}[shift={(0,0)}] % \Delta X
        \draw (1,0) -- ++(1,0) -- ++(0,1);
        \draw (0,1) -- ++(0,1) -- ++(1,0);
        \draw[fill=gray!30] (0,0) -- ++(1,0) -- ++(0,1) -- ++(-1,0) -- ++(0,-1);
        \draw[fill=gray!30] (1,1) -- ++(1,0) -- ++(0,1) -- ++(-1,0) -- ++(0,-1);
        \node at (1,-0.25) {$\Delta X$};
    \end{scope}
    \begin{scope}[shift={(2.5,1)}] % -- \psi -->
        \draw[->] (0,0) -- (1,0);
        \node at (0.5, 0.25) {$\psi$};
    \end{scope}
    \begin{scope}[shift={(4,0)}] % \psi(\Delta X)
        \draw (1,0) -- ++(1,0) -- ++(0,1);
        \draw (0,1) -- ++(0,1) -- ++(1,0);
        \draw[fill=gray!30] (0,0) -- ++(1,0) -- ++(0,1) -- ++(-1,0) -- ++(0,-1);
        \draw[fill=gray!30] (1,1) -- ++(1,0) -- ++(0,1) -- ++(-1,0) -- ++(0,-1);
        \draw (0, 1.5) -- ++(1,0);
        \draw (0.5, 1) -- ++(0,1);
        \draw (1, 0.5) -- ++(1,0);
        \draw (1.5, 0) -- ++(0,1);
        \draw[gray,dashed,thick] (0, 0.5) -- ++(1,0);
        \draw[gray,dashed,thick] (0.5, 0) -- ++(0,1);
        \draw[gray,dashed,thick] (1, 1.5) -- ++(1,0);
        \draw[gray,dashed,thick] (1.5, 1) -- ++(0,1);
        \node at ( 1,-0.25) {$\psi(\Delta X)$};
    \end{scope}
    \begin{scope}[shift={(6.5, 1)}]
        \node at (0,0) {$\subset$};
    \end{scope}
    
    \begin{scope}[shift={(0,-3.25)}] % \Delta\Omega
        \draw (1,0) -- ++(1,0) -- ++(0,1);
        \draw (0,1) -- ++(0,1) -- ++(1,0);
        \draw (0,0) -- ++(1,0) -- ++(0,1) -- ++(-1,0) -- ++(0,-1);
        \draw (1,1) -- ++(1,0) -- ++(0,1) -- ++(-1,0) -- ++(0,-1);
        \node at (1,-0.25) {$\Delta \Omega$};
    \end{scope}
    \begin{scope}[shift={(2.5,-2.25)}] % -- \psi --> to omega
        \draw[->] (0,0) -- (2.5,0);
        \node at (1.25, 0.25) {$\psi$};
    \end{scope}
    \begin{scope}[shift={(7,0)}] % \Delta Y
        \draw (1,0) -- ++(1,0) -- ++(0,1);
        \draw (0,1) -- ++(0,1) -- ++(1,0);
        \draw[fill=gray!30] (0,0) -- ++(1,0) -- ++(0,1) -- ++(-1,0) -- ++(0,-1);
        \draw[fill=gray!30] (1,1) -- ++(1,0) -- ++(0,1) -- ++(-1,0) -- ++(0,-1);
        \draw (0, 1.5) -- ++(1,0);
        \draw (0.5, 1) -- ++(0,1);
        \draw (1, 0.5) -- ++(1,0);
        \draw (1.5, 0) -- ++(0,1);
        \draw (0, 0.5) -- ++(1,0);
        \draw (0.5, 0) -- ++(0,1);
        \draw (1, 1.5) -- ++(1,0);
        \draw (1.5, 1) -- ++(0,1);
        \node at (1,-0.25) {$\Delta Y$};
    \end{scope}
        \begin{scope}[shift={(5.5,-3.25)}] % \psi(\Delta \Omega)
        \draw (1,0) -- ++(1,0) -- ++(0,1);
        \draw (0,1) -- ++(0,1) -- ++(1,0);
        \draw (0,0) -- ++(1,0) -- ++(0,1) -- ++(-1,0) -- ++(0,-1);
        \draw (1,1) -- ++(1,0) -- ++(0,1) -- ++(-1,0) -- ++(0,-1);
        \draw (0, 1.5) -- ++(1,0);
        \draw (0.5, 1) -- ++(0,1);
        \draw (1, 0.5) -- ++(1,0);
        \draw (1.5, 0) -- ++(0,1);
        \draw (0, 0.5) -- ++(1,0);
        \draw (0.5, 0) -- ++(0,1);
        \draw (1, 1.5) -- ++(1,0);
        \draw (1.5, 1) -- ++(0,1);
        \node at (1,-0.25) {$\Delta \phi(\Omega)$};
    \end{scope}

    \node[rotate=-45] at (5.25, -0.8) {$\subset$};
    \node[rotate=225] at (7.75, -0.8) {$\subset$};    
\end{tikzpicture}
\caption{In the scenario above, $\Omega$ has sufficient resolution to describe $\Delta X$ but not $\Delta Y$. If we refine $\Omega$ to $\Omega'$, we are able to describe both $\Delta X$ with $\psi(\Delta X)$ and also $\Delta Y$. The grey colouring indicates equiprobable distribution of outcomes. The dashed lines show that there are boundaries in $Y$ which are not seen in $X$: $Y$ is strictly more informative than $X$, and they are both less informative than the measure of the entire space, $\Delta \phi(\Omega)$. $\Delta X$ and $\psi(\Delta X)$ are equivalent and have equal measure. $\Delta X$ and $\Delta Y$ correspond to distinct elements in $\delta \Omega$, as they are not equivalent in the finer space $\Omega'$.}
\label{FIGURE_Inclusions}
\end{figure}

This invariance under refinement captures the truly interesting structure in $\Delta \Omega$. As we are now able to refine $\Omega$, we can in principle refine it indefinitely to construct finer and finer spaces in which to decompose, capturing partitions at all levels as we go. In the next section we extend this construction to explore potential descriptions of continuous variables using the logarithmic decomposition.

\end{example}

\ifSubfilesClassLoaded{\bibliographystyle{plain} \bibliography{main} }{}
\end{document}

\section{Continuous Logarithmic Decomposition}
\label{SECTION_Continuous}

In the previous section we constructed the space $\delta \Omega$ for exploring equivalence classes of logarithmically decomposable quantities under refinements of the outcome space. In this section we will explore how we can use a limiting process inside of $\delta \Omega$ to approximate continuous variables, in a scenario analogous to the completion of the real numbers.

To define the `closeness' of an approximation to a continuous variable, we shall require that our approximation uniformly converges to a continuous variable.

% I MIGHT TRY REWRITING THIS DEFINTION
\begin{definition}
Let $p_X$ be the probability density of a continuous random variable $X$ on some continuous outcome space $\Omega$. Let $(X_n)_{n\in \N}$ be a sequence of discrete variables whose outcomes represent distinct subsets $E_{n,m}$ of $\Omega$, where $m$ is indexing the different events in $X_n$, such that, given any event $E_{n,m} \subseteq \Omega$,

\begin{equation}
P(X_n = E_{n,m}) = \int_{E_{n, m}} p_X \,\mathrm{d}x
\end{equation}

That is, for an outcome in the discrete variable $X_n$, there is a corresponding subset in the continuous space $\Omega$, over which we can integrate the continuous probability density to find the discrete probability. At each stage, $X_n$ is breaking the space $\Omega$ into $m$ pieces.

We will say that the sequence $X_n$ \textbf{uniformly converges to $p_X$} if the discrete probability density
\begin{equation}
p_n(\omega) = \begin{cases} \frac{1}{P(X_n = E_{n,m})} & \omega \in E_{n,m}
\end{cases}
\end{equation}
uniformly converges to $p_X(\omega)$. That is,
\begin{equation}
\forall \epsilon > 0, \exists N \in \N\, [n \geq N \implies \forall\omega [ |p_n(\omega) - p_X(\omega)| < \epsilon ] ].
\end{equation}

\end{definition}

This definition captures the idea that $X_n$ approximates $p_X$ in a limiting process, by considering the probability measure integrated over each region in $X_n$. 

\begin{definition}
Now suppose that we have a sequence $X_n$ of finite variables which is \textbf{uniformly convergent} to $p_X$ over an outcome space $\Omega$ as above, but where for each $n$,
\begin{equation}
P(X_n = E_{n,m_1}) = P(X_n = E_{n, m_2})
\end{equation}
for every $m_1, m_2$, and we require further that this probability is tending to zero as $n$ increases. That is, all events have equal probability at each step of the refinement, and $\Omega$ is partitioned into gradually smaller and smaller pieces. We shall say that such a system of variables is \textbf{uniformly and equiprobably convergent} to $p_X$.
\end{definition}

This definition represents a maximum-entropy based characterisation of the continuous distribution; it represents the status quo of our knowledge with optimal coding.

Coupled with our construction of the space $\delta \Omega$, we will now \textbf{complete} the space so that we can represent continuous variables in our measure-theoretical perspective.

\begin{definition}
Let $X_n$ and $Y_n$ be two sequences of discrete variables on $\Omega$. We shall write $X_n \sim Y_n$ and call them \textbf{equivalent} if they are both uniformly and equiprobably convergent to $p_X$.
\end{definition}

\begin{definition}
Let $p_X$ and $p_M$ be continuous probability distributions on the space $\Omega$. We select a representative $M_n$ of the class of sequences of variables uniformly and equiprobably convergent to $p_M$ with the additional property that $M_{n+k}$ is a refinement of $M_n$ for all $k \in \N$. Let $X_n$ be a new random variable defined on the same outcome space $\Omega$, with the same events as $M_n$, namely $E_{n, 1}, E_{n, 2}, \ldots$, but with probability distribution given by
\begin{equation}
P(X_n = E_{n, m}) = \int_{E_{n,m}} p_X \,\mathrm{d}x.
\end{equation}
That is, $X_n$ is a sister variable to $M_n$, which is defined on the same events but with \textit{different} probabilities. We will refer to $M_n$ either as an \textbf{invariant measure}, to follow Jaynes' alteration to the differential entropy \cite{jaynes1957information, jaynes1968prior} or we may simply refer to $M_n$ as a reference or prior measure. We shall refer to $X_n$ simply as a variable constructed over the measure $M_n$.
\end{definition}

This setup should feel familiar to that of the Kullback-Leibler divergence \cite{kullback1951information}. In essence we have a partition of $\Omega$ which is optimised for a code based on $p_M$, and we are considering an alternative distribution $p_X$. It should come as no suprise then that we have the following result.

\begin{proposition}
\label{PROPOSITION_KullbackLeibler}
Let $M_n$ be uniformly and equiprobably convergent to $p_X$ and $X_n$ correspond to the same events in $\Omega$ as above. Then for all $n$ we have
\begin{equation}
-D_{\mathrm{KL}}(X_n \,||\, M_n) = \mu(\Delta X_n) - \mu(\Delta M_n),
\end{equation}
and in the limit, the Kullback-Leibler divergence is given by
\begin{equation}
-D_{\mathrm{KL}}(p_X \,||\, p_M) = \lim_{n \to \infty} \left[ \mu(\Delta X_n) - \mu(\Delta M_n) \right].
\end{equation}
\end{proposition}
In this scenario, our variables match the behaviour of the limiting density of discrete points of Jaynes \cite{jaynes1968prior, jaynes1957information}. As a result, this gives us a measure which is equal to the negative Kullback-Leibler divergence from $M$ to $X$. It does not appear to hold for an arbitrary choice of partition given the \textbf{invariant measure} $p_M$. This appears to be due to the fact that arbitrary partitions would represent non-optimal coding of $p_M$.

It is unsurprising that when considering single variables in isolation that the measure of the set $\Delta X_n$ ceases to be finite as we approximate a continuous variable. This is the natural behaviour for the natural limit of the discrete entropy. As usual, however, certain classes of sets, while appearing to become infinitely refined, do have stable measures -- as we explain next.

\subsection{Convergent measures under refinement}

As per the usual scenario, quantities such as \textbf{mutual information} ($\Delta X \cap \Delta Y$) and \textbf{co-information} ($\Delta X \cap \Delta Y \cap \Delta Z$ and above) remain finite in measure as they approximate a given distribution, even as the marginal entropies do not. The logarithmic decomposition approach provides an interesting perspective on why this is the case.

We give a brief example to illustrate this property.

\begin{example}

Suppose we have an entirely redundant system, where one bit of information is shared between two variables $X$ and $Y$. In our current setup, we need only consider the outcome space $\Omega = \{00, 01, 10, 11\}$ to capture all of the behaviour of the system. In this scenario, $p(00) = p(11) = 0.5$, and $p(01) = p(10) = 0$.

Given two parts $P_1$ and $P_2$ of a partition, we will use the notation $P_1 * P_2$ to denote all of those atoms in $\Delta (P_1 \cup P_2)$ which strictly lie across the boundary between $P_1$ and $P_2$\footnote{While this roughly captures a similar idea as $\Delta$, we'll avoid using that notation as we have almost exclusively used $\Delta$ for variables thus far.}.

The mutual information given by this system is provided exclusively by the $\{00, 11\}$ atom, so considering $\Omega$ restricted to $\{00, 11\}$ is sufficient to capture all of the behaviour.

Suppose now that we were to refine our space somewhat further, so that we have four outcomes for $X$ and $Y$, which we call $a, b, c, d$. Then the new, refined outcome space is given by $\Omega' = \{aa, ab, ac, ad, ba, bb, bc, bd, ca, cb, cc, cd, da, db, dc, dd\}$.

\begin{figure}[h]
    \centering
\begin{tikzpicture}[scale=0.9]
    \begin{scope}[shift={(0,0)}] % \Delta X and \Delta Y.
        \draw (1,0) -- ++(1,0) -- ++(0,1);
        \draw (0,1) -- ++(0,1) -- ++(1,0);
        \draw[fill=lightgray] (0,0) -- ++(1,0) -- ++(0,1) -- ++(-1,0) -- ++(0,-1);
        \draw[fill=lightgray] (1,1) -- ++(1,0) -- ++(0,1) -- ++(-1,0) -- ++(0,-1);
        \node at (0.5, -0.25) {$0$};
        \node at (1.5, -0.25) {$1$};
        \node at (-0.25, 0.5) {$0$};
        \node at (-0.25, 1.5) {$1$};
        \node at (1,-0.75) {$X$};
        \node at (-0.75, 1) {$Y$};
        \draw[->] (2.65,1) -- ++(1,0);
    \end{scope}
    \begin{scope}[shift={(5, 0)}]
        \draw (1,0) -- ++(1,0) -- ++(0,1);
        \draw (0,1) -- ++(0,1) -- ++(1,0);
        \draw[fill=lightgray] (0,0) -- ++(1,0) -- ++(0,1) -- ++(-1,0) -- ++(0,-1);
        \draw[fill=lightgray] (1,1) -- ++(1,0) -- ++(0,1) -- ++(-1,0) -- ++(0,-1);
        \draw (0.5, 0) -- (0.5, 1);
        \draw (0, 0.5) -- (1, 0.5);
        \draw (1.5, 1) -- ++(0,1);
        \draw (1, 1.5) -- ++(1,0);
        \node at (0.25, -0.25) {$a$};
        \node at (0.75, -0.25) {$b$};
        \node at (1.25, -0.25) {$c$};
        \node at (1.75, -0.25) {$d$};
        \node at (-0.25, 0.25) {$a$};
        \node at (-0.25, 0.75) {$b$};
        \node at (-0.25, 1.25) {$c$};
        \node at (-0.25, 1.75) {$d$};
        \node at (1,-0.75) {$X'$};
        \node at (-0.75, 1) {$Y'$};
    \end{scope}
\end{tikzpicture}
\caption{Given two continuous variables $X$ and $Y$, their mutual information $I(X;Y) = \mu(\Delta X \cap \Delta Y)$ is convergent under refinements.}
\end{figure}
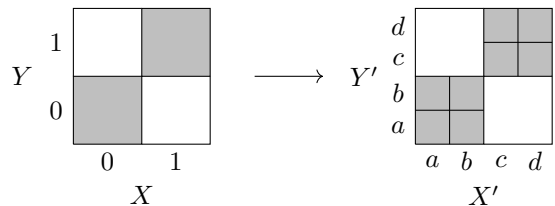

Let us label the variables after the refinement with $X'$ and $Y'$. As a consequence of the refinement, we see that 
\begin{equation}
\{00\}*\{11\} \mapsto \{aa, ab, ba, bb\}*\{cc, cd, dc, dd\}
\end{equation}
and further, by corollary \ref{COROLLARY_AllRefinementsStable}, we know that
\begin{equation}
\mu(\{00\}*\{11\}) = \mu(\{aa, ab, ba, bb\} * \{cc, cd, dc, dd\}).
\end{equation}
Now, we have the convenient fact that
\begin{align}
\Delta X \cap \Delta Y \sim \,\, &(\Delta X' \cap \Delta Y')|_{\{aa, ab, ba, bb\}} \notag \\ &\cup (\Delta X' \cap \Delta Y')|_{\{cc, cd, dc, dd\}} \\ & \cup \{aa, ab, ba, bb\} * \{cc, cd, dc, dd\}. \notag
\end{align}
In this scenario, the $\{aa, ab, ba, bb\}*\{cc, cd, dc, dd\}$ term is carrying all of the original entropy of the system before refinement. The other two terms are newly provided by the refinement.

More intuitively speaking, this three-part decomposition says simply that the mutual information between $X$ and $Y$ is given by atoms which either lie completely in the top right, completely in the bottom left, or straddle both (those straddling both sum to the original mutual information). Speaking more abstractly, the mutual information in this case corresponds to two local interactions and one global interaction.

These sets are all disjoint, so the measure is additive. But notice now that $\mu(\Delta \{aa, ab, ba, bb\}) = \mu(\Delta \{cc, cd, dc, dd\}) = 0$, because, looking at these smaller systems in their own right, their contribution to the entropy looks like the mutual information shared between two independent binary variables (even if the probabilities in this case only sum to one half in each system). As a result, they both cancel, giving zero. Hence we have that
\begin{equation}
\mu(\Delta X \cap \Delta Y) = \mu(\Delta X' \cap \Delta Y').
\end{equation}
That is, our `global' interaction (atoms crossing the diagonal boundary, which are the image of the interaction present before refinement) is left intact at 1 bit, and the `local' interactions (newly introduced with refinement) perish.  As such, as we approximate a smooth probability density, the local interactions (those added at each level) will become very small as the neighbourhood becomes increasingly small, as the local systems look increasingly uniform, causing a finite return on mutual information under refinement.
\end{example}

\begin{definition}
\label{DEFINITION_MicroMacro}
We refer to this property, whereby entropy is the sum of local, microscopic interactions, and global, macroscopic interactions (or even interactions at all scales), the \textbf{micro-macro principle}.

In other words, given subsystems $S \subseteq \Omega$ indexed by $\mathcal{S}$ which partition $\Omega$, entropy consists of contributions inside of subsystems $\bigcup_{S \in \mathcal{S}} \Delta S$ and between subsystems $\Star_{P\subseteq \mathcal{S}} P$.
\end{definition}

We have shown in this section that our set-theoretic perspective on entropy is not limited to the discussion of discrete variables, even if the intuition is far clearer in this case. We demonstrated how the measure interacts with refinements and how this can be applied in sequences to construct measures for continuous variables. Lastly, we saw that such a decomposition gives a nice perspective on the finiteness of mutual information and an interesting macroscopic way of separating contributions to entropy.

\ifSubfilesClassLoaded{\bibliographystyle{plain} \bibliography{main} }{}
\end{document}

\section{The Dyadic and Triadic Systems}
\label{SECTION_TriadicDyadic}

To further demonstrate the utility of the logarithmic decomposition described, we apply the decomposition to two systems initially considered together by James and Crutchfield in 2016 \cite{james2017multivariate}. These two systems are constructed so as to have identical conditional entropies and co-information, rendering them indistinguishable when using classical techniques. \revlbl{R1.1, R1.2}\added{James and Crutchfield even go so far as to say that `no standard Shannon-like information measure, and exceedingly few nonstandard methods, can distinguish the two' \cite{james2017multivariate}. As the main result and culmination of this paper, we demonstrate here that the logarithmic decomposition, which parameterises all possible classical information measures on a given outcome space $\Omega$, allows us to distinguish between the Dyadic and Triadic system structures (without the use of partial information decomposition (PID) methods or similar \cite{bertschinger2014quantifying,ince2017measuring,williams2010nonnegative,kolchinsky2022novel}.}

\begin{figure}[ht]
\begin{center}
    \includegraphics[scale = 0.1]{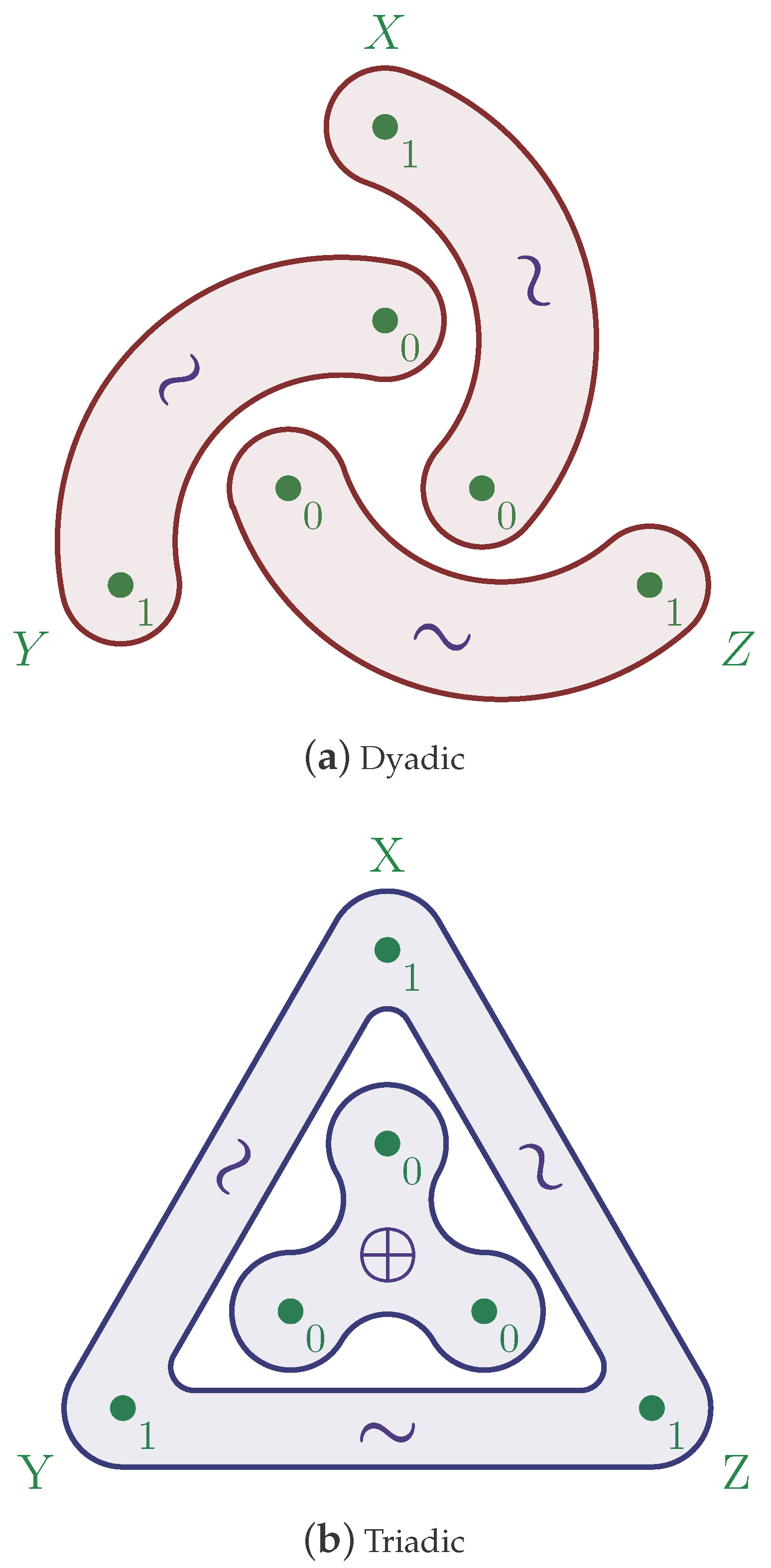}
    \caption{The (a) Dyadic and (b) Triadic systems, taken from the original paper by James and Crutchfield \cite{james2017multivariate}. In this set-up, all variables are represented with two binary symbols. The tilde $\sim$ represents coupled bits; these bits always observe the same symbol. The $\oplus$ represents an XOR gate, where $Z = \text{XOR}(X,Y)$. In the dyadic system, it is expected that there is no shared information and no synergy. In the triadic system, there is one bit of sharedness and one bit of synergy, which cancel each other out in the co-information.}
\end{center}
\end{figure}

The dyadic system consists of three coupled bits, distributed pairwise between each of three variables. In this case, it is expected that there is no information shared between the three variables (in the sense of a redundancy function for a partial information decomposition -- see \cite{williams2010nonnegative}, for example). This is accurately reflected by the fact that the co-information between all three variables is precisely zero.

The triadic system, on the other hand, is constructed from one bit, coupled between three variables, and one XOR gate. The coupled bit should contribute 1 bit of entropy to the co-information, but the XOR gate is thought to remove 1 bit of entropy from the co-information, again leaving this fixed at zero.

These two systems have the intriguing property that their co-information structures are completely identical, and yet they have explicitly distinct characteristics. James and Crutchfield note that ``no standard Shannon-like information measure, and exceedingly few nonstandard methods, can distinguish the two" \cite{james2017multivariate}.

\begin{figure}[ht]
\begin{center}
\begin{tikzpicture}
    \newcommand{\growth}{0.8}
    
  % Circles
  \draw (0*\growth,0*\growth) circle (2);
  \draw (2*\growth,0*\growth) circle (2);
  \draw (1*\growth,1.732*\growth) circle (2);
  
  % Label in the middle
  \node[scale=1.5] at (1*\growth,0.5*\growth) {$0$};
  \node[scale=1.5] at (1*\growth,0.5*\growth - 1.5) {$1$};
  \node[scale=1.5] at (1*\growth + 1.5*0.866 ,0.5*\growth + 1.5*0.5) {$1$};
  \node[scale=1.5] at (1*\growth - 1.5*0.866 ,0.5*\growth + 1.5*0.5) {$1$};
  
  \node at (-2*\growth-0.1, -2*\growth-0.1) {$X$};
  \node at (4*\growth+0.1, -2*\growth-0.1) {$Y$};
  \node at (1*\growth, 4*\growth + 0.6) {$Z$};
\end{tikzpicture}
\caption{The $I$-measure applied to both the Dyadic and Triadic systems gives the same distribution of conditional entropies, despite their distinct qualitative structures.}
\end{center}
\end{figure}
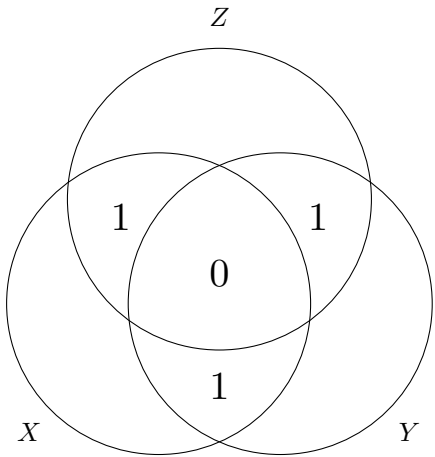

Using our logarithmic decomposition we can separate the structure of these two systems. To explain how, we give a definition.

\begin{definition}
Let $C$ be a set of logarithmic atoms. We use the notation
\begin{equation}
R_n(C) = \{c \in C: \exists\, c' \in C, \deg(c') = n \text{ such that } c' \subseteq c \}.
\end{equation}
That is, $R_n(C)$ consists of all of the atoms which, as a set, contain another atom of degree $n$ inside the set. We can also think of this structure as reflecting elements which lie inside of the \textbf{upper set} generated by degree $n$ atoms inside of $C$ in the partial order given by inclusion.
\end{definition}

We note that the definition of $R_n$ is completely symmetric in that it makes no conventions about labelling -- it depends only on the underlying structure of the set $C$.

\begin{theorem}
The dyadic and triadic systems have distinct structures under the logarithmic decomposition.
\end{theorem}
\begin{proof}
We have that
\begin{equation}
\mu(R_2(\Delta X_{\textbf{dy}} \cap \Delta Y_{\textbf{dy}} \cap \Delta Z_{\textbf{dy}})) = 0
\end{equation}
whereas
\begin{equation}
\mu(R_2(\Delta X_{\textbf{tri}} \cap \Delta Y_{\textbf{tri}} \cap \Delta Z_{\textbf{tri}})) = 1.
\end{equation}
\end{proof}

By virtue of proposition \ref{PROPOSITION_EntropyExpressions} \added{and in keeping with remark \ref{REMARK_Measure_everything}}, it can be seen that the logarithmic decomposition corresponds to the $I$-measure decomposed over the set of all partitions of the outcome space $\Omega$. Further to this, every single atom and combination of atoms has a corresponding entropy expression. For this reason, the decomposition is \replaced{fundamentally}{essentially} classical, with the helpful property that it is still able to structurally discern between the dyadic and triadic systems. We believe it might suffice as \replaced{one potential}{the intended} discriminatory measure \added{as} discussed by James and Crutchfield~\cite{james2017multivariate}.

\ifSubfilesClassLoaded{\bibliographystyle{plain} \bibliography{main} }{}
\end{document}

\section{Conclusion}

\subsection{Main Contributions}
In this work we developed a signed measure space which refines the prevailing $I$-measure of Yeung \cite{yeung1991new} to produce a significantly refined signed measure space for Shannon entropy. We demonstrated that this space is consistent with \added{and finer than} the $I$-measure and can be used to express many information-theoretic quantities, including the mutual information and co-information, along with quantities exhibiting multiplicities such as the O-information \cite{rosas2019quantifying}, total correlation \cite{watanabe1960information} and dual total correlation \cite{te1978nonnegative}. Further to this, we also showed that the decomposition can express other quantities which were previously inexpressible using the $I$-measure alone, such as the G\'acs-K\"orner common information \added{\cite{gacs1973common}}, minimally sufficient statistic (MSS) \cite{fisher1922mathematical, lehmann2011completeness} and \added{the functional common information of James and Crutchfield \cite{james2017multivariate}}.

We constructed the measure $\mu$ by first constructing an intermediate measure we referred to as `loss', which captures the information lost when merging outcomes \added{\cite{baez2011characterization}}. This choice is quite natural and allowed us to move from a variable-scale language of entropy to an outcome-scale language of entropy, giving a strong foundation for a qualitative theory of information. This perspective has a pleasing naturality to it, in that the operational interpretation of the loss is very much clear and scales homogeneously, both classically with degree 1 and with degree $d$ for the $d$-th Tsallis entropy \cite{baez2011characterization}.

We then applied a M\"obius inversion on the loss over the lattice of all subsets of the outcome space $\Omega$ to construct the measure $\mu$, which, when defined on finite outcome spaces, was shown to come naturally equipped with many intriguing and useful properties which are lost at coarser granularities. For example, we saw that each logarithmic atom $b \in \Delta\Omega$ has a fixed signs depending only on its \textit{degree} - the number of outcomes to which it relates (see theorem \ref{AlternatingDerivatives}), and we also saw that the magnitude of entropy contributions from atoms monotonically decreases with increasing degree (see corollary \ref{COROLLARY_MuMagnitude}). Constructing these atoms also allowed us to resolve the discrepancy between coding and shared information; coded information can only be represented by a variable when it coincides with certain classes of collections of atoms, while mutual and co-information are not necessarily representable in the same way, providing unique insight as to why ``common information is much less than mutual information'' \cite{gacs1973common}.

More than this, we saw that atoms correspond to pieces which capture different qualitative aspects of conferred information -- all atoms have an operational meaning in that they are present when a variable \textbf{can observe a change} in some subset of $\Omega$. As such, this framework provides a transition between the quantity-led approach of classical information theory, to the \textit{quality-and-quantity}-based description of a signed measure space. In such a space, the subsets of the space (consisting of groups of atoms) correspond to qualitative knowledge about outcomes, and the measure provides a quantitative metric to find their contributions to the entropy. We provided a definition for \textbf{logarithmically decomposable} quantities where this set-theoretic representation can be utilised to its full potential.

We explored in section \ref{SECTION_Inclusions} how the decomposition interacts with refinements of the outcome space $\Omega$ and applied this to the study of continuous variables in section \ref{SECTION_Continuous}. In this case, we recovered the limiting density of discrete points of Jaynes from our set-theoretic perspective \cite{jaynes1957information}. Moreover, we found that the finiteness of mutual information in the continuous case follows from a novel cancellation argument, illuminated by a set-theoretic decomposition into \textbf{microscopic} and \textbf{macroscopic} pieces.

Finally, \added{as the main result of this work}, we applied all of our qualitative methods to the Dyadic and Triadic systems as presented by James and Crutchfield \cite{james2017multivariate}, showing that, using only the \added{qualitative behaviour} described by our decomposition, \added{it is possible} to discern between the two systems using an argument based on pairwise contribution to entropy (our quantity $\mu(R_2)$); something which has, classically, not been previously seen. \revlbl{R1.1, R1.2}\added{This surprising result challenges the prevailing belief that the qualitative separation of systems requires an extension to classical information theory (usually presented as Partial Information Decomposition, PID \cite{bertschinger2014quantifying, ince2017measuring, kolchinsky2022novel, williams2010nonnegative}).}

\subsection{Limitations}

The logarithmic decomposition given in this work does come with large computational requirements if one is unwilling to make clever counting arguments. We note that in the general case the total number of atoms grows with $2^{|\Omega|} - |\Omega| - 1$. Keeping track of the value of each of these atoms proves to be computationally challenging when scaling with large systems, but there are alternative routes for calculating quantities of interest. We believe, for example, there might be a simplified representation of the subset $R_n$ as defined in section \ref{SECTION_TriadicDyadic}.

It has been well noted in the literature that Shannon entropy exhibits much algebraic behaviour when viewed from different perspectives. It has, for example, a characterisation in terms of homology \cite{baudot2015homological, vigneaux2017information}, among other perspectives. While we focused only on a few algebraic properties of $\Delta\Omega$ as a lattice here (as appears very frequently in current work on shared information \cite{mediano2021towards, williams2010nonnegative, bell2003co, shannon1953lattice}), there may be other algebraic properties of $\Delta \Omega$ that warrant investigation. It has not escaped our notice, for example, that the refinement operation of definition \ref{DEFINITION_ThreeOperators} could perhaps be better viewed as a ring homomorphism on $\Z\Delta\Omega$.

While we noted that the Tsallis entropy loss has a natural homogeneity property (as seen in \cite{baez2011characterization}), we did not explore how our event-based decomposition works when applied to these generalised entropies. In particular, it is unclear whether or not lemmas \ref{LEMMA_InteriorLossIdentity} and \ref{LEMMA_MuAtInfinity} have corresponding results for general Tsallis entropies.

\added{While the logarithmic decomposition can represent many quantities \textit{without refinement} as demonstrated in section \ref{SECTION_Quantities}, there are quantities where a refinement is necessary and has to be carefully taken into account. The Wyner common information \cite{wyner1975common, xu2013wyner} is one such quantity. While a refinement is necessary, this does not mean that the decomposition provided here does not offer the potential for additional insight. In particular, we have left the notion of independence essentially unexplored. The careful reader will note that for independent variables $X$ and $Y$, it is not generally the case that $\Delta X\cap \Delta Y = \varnothing$. Instead, one finds a collection of atoms which appear to meet in \textit{syzygies}, with a total measure of zero. Interrogating the precise structure of these collections of atoms might prove fruitful to widening the reach of this decomposition, but this is beyond the scope of the current work.}

\subsection{Implications}

We foresee that this qualitative language will have much use in dissecting information processing in complex systems, where information quality has been previously difficult to access. Understanding qualitative information processing in the brain, for example, would provide a natural language for understanding neural representations in cognitive and computational neuroscience \cite{luppi2022synergistic, proca2024synergistic, luppi2024synergistic, gelens2024distributed}.

The development of explainable AI might also benefit from a qualitative approach to information theory. The representations of machine learning models are often opaque and difficult to interpret. Understanding qualitative information processing these systems might have significant safety and bias implications for the technology \cite{angelov2021explainable, zhang2021survey, wollstadt2023rigorous}.

The original motivation for the decomposition described here was to enable further development of the partial information decomposition (PID) methodology, which aims to decompose information into representations as \textbf{redundant}, \textbf{unique}, and \textbf{synergistic} information \cite{williams2010nonnegative, bertschinger2014quantifying, rosas2020operational, kolchinsky2022novel, mediano2021towards}. Many versions of the partial information decomposition now exist, though none yet has been conclusively accepted as \textit{the correct} method. Given that the sign of co-information measures is closely tied to the study of redundant and synergistic behaviour in information structure, we expect that the fixed-sign language of the logarithmic decomposition will provide a new perspective for exploring the partial information decomposition problem. As our decomposition allows us to parameterise classical entropy quantities, it might also be possible to either construct a classical PID or show that no such construction can exist -- both of these outcomes would be a significant development in the theory of PID. \revlbl{R1.1, R1.2}\added{The Dyadic and Triadic systems, studied here in section \ref{SECTION_TriadicDyadic}, are also of particular interest to the practitioners of PID.}

\revlbl{R1.2}\added{In a sequel to the current work, the authors demonstrated that this construction, coupled with an algebraic interrogation of the structure of the logarithmic decomposition, offers novel techniques for studying the boundedness of various information quantities. In the sequel, it was shown that the purely-synergistic behaviour of the XOR gate (another beloved case-study in the practice of PID) is, in fact, unique \cite{down2025algebraic}. More than this, the framework of the decomposition allows this to be demonstrated \textit{purely algebraically}.}

\subsection{Summary}

The structure of the decomposition given in this work is remarkably rich, providing new perspectives on the nature of coded information \added{and the self-similar nature of entropy}. We demonstrated that our decomposition is endowed with many properties that coarser measures such as the $I$-measure do not have, and it can be used to describe many quantities in a set-theoretical fashion. We expect that this new language, coupled with a rigorous interrogation of the algebraic structure of this decomposition, will provide paths for new perspectives on old bounding problems and an improved understanding of \added{the structure of entropy and shared information}. \revlbl{E2.2}\added{More than this, we believe that the logarithmic decomposition introduced here will prove to be an exceptionally powerful tool for the rigorous interrogation of Partial Information Decomposition, offering new lines of inquiry in this particularly troubled research area}.

\ifSubfilesClassLoaded{\bibliographystyle{plain} \bibliography{main} }{}

\section*{Acknowledgements}

The authors would like to thank Dan Bor, Fernando Rosas, Robin Ince and Abel Jansma for their input on future directions for this work. %\added{We also thank the anonymous reviewers, for whom this work has been significantly improved - especially in its treatment of the Wyner common information, which was previously erroneous.}

\end{document}

\bibliographystyle{plain}
\bibliography{main.bib}

\clearpage
\appendices

\section{Measures on Simplices}
\label{SECTION_Simplex}

In the interest of justifying that the construction is, in fact, the unique way of naturally refining the $I$-measure, we consider also what the construction would look like for alternative choices of `base space'. What should it mean if the measure cannot be attached to a simplex, for example, and instead requires some alternative backbone? Can we extract the logarithmic decomposition under alternative circumstances?

If we are to suppose that the inclusion-exclusion principle should hold when studying information (and hence that the measure-theoretic perspective is even well-founded at all, which is often in dispute \cite{kolchinsky2022novel}), then an outcome-based language for the entropy should always be viewable in terms of each outcome and interactions between those outcomes:
\begin{align}
H(p_1, p_2, p_3) & = f_1(p_1) + f_2(p_2) + f_3(p_3) \notag \\ & + f_{12}(p_1, p_2) + f_{13}(p_1, p_3) + f_{23}(p_2,p_3) \\ & + f_{123}(p_1, p_2, p_3). \notag
\end{align}
Any function on three variables could, in principle, be separated into additive parts depending only on subsets of those variables on which it depends. That is, given an alternative backbone defined on the probabilities themselves (as must be the case, otherwise the measure is hardly outcome-wise), we can always reduce the situation to studying measures on a simplex.

We can perhaps even argue more than this. Given that all of the knowledge we have about a variable is described by precisely its outcomes and their probabilities, and that, up to introducing more variables and studying unknown interactions, outcomes are equivalent to their probabilities, any measure for information that captures all of this knowledge successfully, and not more \textit{must only} depend on these probabilities.

As such, the construction of this simplex measure appears to always be possible. It may be that in an alternative guise, various components of these atoms $f_i$ co-appear. However, as we have seen, it is always possible to construct information quantities that separate and filter these elements, so for an alternative formulation to be successful, it must at the very least offer some method of computing each atom individually, else it fails to construct all classical information measures on a finite outcome space $\Omega$.

What about refinements on the simplex? What of systems more complex than the simplex, of whom certain components can be taken to represent the simplicial measure? In these cases, too, it is perhaps possible to argue that, as no classical measures can now discern between items finer than those on the simplex, that the additional detail is possibly unnecessary.

In this sense, the decomposition presented in this paper is the signed measure space of entropy that is sufficiently fine, and not finer than what is required, to successfully derive all of the classical information quantities.

\subsection{Dependency}

Mathematically deriving the dependencies of these parts might be indirectly accessible. Given some expression $g(x_1,\ldots, x_n)$ which depends explicitly on $x_1,\ldots, x_n$, we might extract those parts which depend on `at least $x_i$' with
\begin{equation}
F_{i} = \int \left( \frac{\partial g}{\partial x_i} \right) \,\mathrm{d}x_i
\end{equation}
but setting the boundary condition that $F_{1,\ldots, k}(0, \ldots, 0) = 0$. This is equivalent to setting the constant of integration to zero in this case (which might be justified as this shall not depend on $x_i$). This $F_i$ will contain all components depending in any way on $i$, including components explicitly depending on multiple parts. As such this integral corresponds to calculating the quantity
\begin{equation}
\mu(F_i) = \mu (\{\text{all parts depending on $x_i$}\}).
\end{equation}
The general form can be extracted using
\begin{equation}
F_{1,\ldots, k} = \int^{(k)} \left( \frac{\partial^k g}{\partial x_1 \ldots \partial x_k} \right) \,\mathrm{d}x_1 \ldots \mathrm{d}x_k.
\end{equation}
From which the M\"obius inversion formula allows us to extract the required edges and faces of our measure. For example, $f_{1, 2, \ldots, n} = F_{1, 2, \ldots, n}$, while $f_{1, 2, \ldots, n-1} = F_{1, 2, \ldots, n-1} - f_{1, 2, \ldots, n}$, and so on. That is, we can isolate the contributions from a given subset $S$ by considering $F_S$ and removing all contributions from larger sets $R \supset S$, giving us the simplex interiors $f_S$.

All of this is to say, given a decomposition which is \textit{finer} than the logarithmic decomposition (such as the Poisson decomposition proposed by Li \cite{li2023poisson}), we expect it should be possible to extract those atoms in the finer decomposition which correspond as a sum to logarithmic atoms.

\ifSubfilesClassLoaded{\bibliographystyle{plain} \bibliography{main} }{}
\end{document}

\section*{Proofs for results}

\subsection*{Proof of lemma \ref{LEMMA_InteriorLossIdentity}}
\begin{proof}
To simplify we shall also write $f_k = \begin{pmatrix} n-1 \\ k-1 \end{pmatrix}$. This is the number of subsets $S\subseteq \{p_1,\ldots, p_n\}$ of size $k$ which contain a given $p_i$. As we ask for subsets which already contain $p_i$, this is equivalent to asking how many subsets there are of size $k-1$ in $\{p_1,\ldots, p_n\} \setminus \{p_i\}$.

Taking equation \eqref{InclusionExclusion} and using the definition of the total loss function we have
\begin{multline}
    \mu(p_1, \ldots, p_n) \\
= \log \left[  \frac{A_n}{\sigma(p_1) \ldots \sigma(p_n)} \cdot \frac{\sigma(p_1)^{f_{n-1}} \ldots \sigma(p_n)^{f_{n-1}}} {A_{n-1}} \cdots \right. \\ 
\left.\cdots \left( \frac{A_1}{\sigma(p_1)^{f_1} \ldots \sigma(p_n)^{f_1}}\right) ^{(-1)^{n-1}} \right] \\
= \sum_{k = 1}^n {(-1)^{n-k}} \log \left[ \frac{A_k}{\sigma(p_1)^{f_k}\ldots \sigma(p_n)^{f_k}} \right] \\
\end{multline}
Notice that $f_1 = 1$ so that the final term in this sequence with $k=1$ is equal to $\log(1) = 0$. Counting the powers of $\sigma(p_i)$ shows that in the final expression the power of $\sigma(p_i)$ will be $f_n - f_{n-1} + f_{n-2} + \cdots \pm f_2$ (as the $k=1$ term is cancelled by $A_1$). It is a standard result that
\begin{align}
\begin{split}
\label{BinomialStandardResult}
\sum_{k=1}^n (-1)^{(n-k)}f_k &= 0 \quad \text{and hence} \\
\sum_{k=2}^n (-1)^{(n-k)}f_k &= (-1)^{n}
\end{split}
\end{align}
Hence in the final expression the power of $\sigma(p_i)$ is $(-1)^n$. Rewriting $\sigma(p_1)\cdots \sigma(p_n)= A_1$ gives us the result of equation \eqref{ClearerExpression}.
\end{proof}

\subsection*{Proof of lemma \ref{InteriorLossAt0}}
\begin{proof}
We are augmenting $p_1,\ldots, p_n$ with the additional argument $x$, where we will allow $x$ to vary. Let us now write
\begin{equation}
B_k = \prod_{\substack{S \subseteq\{p_1,\ldots, p_n, x\} \\ x \in S \\ |S| = k}} \sigma(S).
\end{equation}
Then equation \eqref{ClearerExpression} becomes
\begin{multline}
\label{BkAndAk}
\mu(p_1,\ldots, p_n, x) = \sum_{k=1}^{n+1} (-1)^{n+1-k} \log(B_k(x))  \\
+  \sum_{k=1}^{n} (-1)^{n+1-k} \log(A_k)
\end{multline}
Here we take $A_k$ to be a product of all terms not containing the argument $x$ as per lemma \ref{InteriorLossAt0}. We notice that the sign of all terms $A_k$ have now flipped, but are otherwise identical. We want to show that as $x\to 0$ that these two sums will cancel. Recall that $B_k(x)$ is a product of terms of the form $\sigma(p_1,\ldots, p_n, x) = (p_1 + \cdots + p_n+ x)^{(p_1 + \ldots + p_n + x)}$ for subsets of size $k$. We see that
\begin{equation}
\lim_{x\to 0} \sigma(p_1,\ldots, p_n, x) = \sigma(p_1,\ldots, p_n)
\end{equation}
By the product and quotient rules for limits, we hence also have that
\begin{equation}
\lim_{x\to 0} B_k = A_{k-1}
\end{equation}
Inserting this into equation \eqref{BkAndAk} we see that both sides immediately cancel to give zero as $x\to 0$.
\end{proof}

\subsection*{Proof of lemma \ref{LEMMA_MuAtInfinity}}

\begin{proof}
Using the expression of lemma \ref{LEMMA_InteriorLossIdentity} and the notation for $B_k(x)$ from lemma \ref{InteriorLossAt0} we can write
\begin{multline}
\mu(p_1,\ldots, p_{n-1}, x) = \sum_{k=1}^n (-1)^{n-k}\log (B_k(x)) + \\
\sum_{k=1}^{n-1} (-1)^{n-k}\log (A_k)
\end{multline}
Where we have omitted the term in $A_n$ because any subset of $\{p_1,\ldots, p_{n-1}, x\}$ of size $n$ is certain to contain $x$. We immediately see that the second expression is equal to $-\mu(p_1,\ldots, p_{n-1})$. It therefore suffices to show that the first expression in the $B_k(x)$ tends to 0 as $x\to \infty$

Writing the logarithm of $B_k(x)$ as a single fraction, we know by the standard binomial result in equation \eqref{BinomialStandardResult} that the number of factors on the top and the bottom of the fraction containing $x$ is equal. Let the number of factors be $m$. Then, expanding the expression in $B_k(x)$, we see it is dominated on the top and the bottom by an $x^m$ term. This term will dominate as $x\to \infty$, so that the fraction tends to 1 and the logarithm in $x$ will tend to 0, leaving us with
\begin{equation}
\lim_{x \to \infty} \mu(p_1,\ldots, p_{n-1}, x) = -\mu(p_1,\ldots, p_{n-1}),
\end{equation}
giving the result immediately.
\end{proof}

\subsection*{Proof of theorem \ref{AlternatingDerivatives}}

\begin{proof}
We will prove this by induction on $n$. To start, we demonstrate that the derivative of $\mu$ has some useful properties. Using standard results and utilising the notation of lemma \ref{LEMMA_InteriorLossIdentity}, we have that
\begin{multline}
\frac{\partial}{\partial x} \, \sigma(x, p_2,\ldots, p_{k}) = \sigma(x, p_2,\ldots,p_{k}) \\ \cdot\left[ \log(x + p_2+\cdots+p_{k}) + 1 \right]
\end{multline}
We restate the identity in equation \eqref{BkAndAk} for $n-1$ fixed probabilities:
\begin{multline}
\mu(p_1,\ldots, p_{n-1}, x) = \sum_{k=1}^{n} (-1)^{n-k} \log(B_k(x))  \\
+ \sum_{k=1}^{n-1} (-1)^{n-k} \log(A_k)
\end{multline}
The second sum does not depend on $x$. Differentiating with respect to $x$ we obtain
\begin{align}
\begin{split}
&\frac{\partial \mu}{\partial x}(p_1,\ldots, p_{n-1}, x) \\
=& \sum_{\substack{S \subseteq \{p_1,\ldots, p_{n-1}, x\} \\ x\in S}} (-1)^{n-|S|}\frac{\partial}{\partial x} \, \log(\sigma(S))\\
=& \sum_{\substack{S \subseteq \{p_1,\ldots, p_{n-1}, x\} \\ x\in S}} (-1)^{n-|S|} \frac{\sigma'(S)}{\sigma(S)} \\
=& \sum_{\substack{S \subseteq \{p_1,\ldots, p_{n-1}, x\} \\ x\in S}} (-1)^{n-|S|} \left[ \log\left( \sum_{s \in S} s \right) + 1 \right] \\
\end{split}
\end{align}
The total number of subsets $S\subseteq \{p_1,\ldots, p_{n-1}\}$ of size $k$ is $\begin{pmatrix} n-1 \\ k \end{pmatrix}$, so by the standard result in equation \eqref{BinomialStandardResult} the $+1$ terms will cancel leaving only an alternating sum of logarithms.

To simplify we shall write
\begin{equation}
E_n(x) = \left( (-1)^{n}\frac{\partial \mu}{\partial x}(x, p_2,\ldots, p_{n}) \right)
\end{equation}
for $n\in \N$. Doing this gives us a sequence $(E_n(x))_{n\in \N}$ removes the alternating factor $(-1)^n$, allowing us to focus on the alternating sign over $m$.

For example
\begin{equation}
E_3(x) = \log \frac{(p_1+x)(p_2+x)}{(p_1+p_2+x)(x)}.
\end{equation}
Note that all of the even subsets will now appear on the top of the fraction and the odd subsets will appear on the bottom. 

For the first case with $n=2$ we have
\begin{align}
\begin{split}
\frac{\partial \mu}{\partial x} (x, p_2) &= E_2(x) \\
&= \log \frac{x+p_2}{x}
\end{split}
\end{align}
which is clearly greater than 0 for all $x \in \R^+$. The successive derivatives of $E_2(x)$ will continue to alternate in sign for $x\in \R^+$ using the standard power rule.

As we also know that $\mu(x, p_2) = L(x, p_2) > 0$, the result holds for $n=2$. We now suppose that the statement is true for $n-1$.

We notice that
\begin{equation}
E_{n}(x) = E_{n-1}(x) - E_{n-1}(x+p_{n})
\end{equation}
Hence
\begin{align}
\label{EQN_difference_of_derivatives}
\begin{split}
& (-1)^n \frac{\partial^m \mu}{\partial x^m} ( x, p_2, \ldots, p_n) \\
&=  \frac{\partial^{m-1}}{\partial x^{m-1}} E_n(x, p_2,\ldots, p_n) \\
&= \frac{\partial^{m-1}}{\partial x^{m-1}} E_{n-1}(x) - \frac{\partial^{m-1}}{\partial x^{m-1}} E_{n-1} (x + p_{n})
\end{split}
\end{align}
However by assumption we have that
\begin{equation}
(-1)^{m-2}\frac{\partial^{m-2}}{\partial x^{m-2}} E_{n-1}(x) > 0
\end{equation}
Hence as the $m-2$-th partial derivative of $E_{n-1}$ has a given sign, we have that the difference between the terms of equation \eqref{EQN_difference_of_derivatives} has the opposite sign. That is,
\begin{equation}
(-1)^{m-1}\frac{\partial^{m-1}}{\partial x^{m-1}} E_n(x, p_2,\ldots, p_n) > 0
\end{equation}
Now, using lemma \ref{InteriorLossAt0} characterizing the interior loss at 0, and using that $E_n$ is strictly positive (negative) for all $x\in \R^+$, the sign of $\mu$ will be strictly negative (positive) for $x\in \R^+$. Hence we have
\begin{equation}
(-1)^n (-1)^m \frac{\partial^{m} \mu}{\partial x^{m}}(x, p_2\ldots, p_n) > 0.
\end{equation}
This completes the inductive argument.
\end{proof}

\subsection*{Proof of corollary \ref{COROLLARY_MuMagnitude}}

\begin{proof}
We saw in lemma \ref{InteriorLossAt0} that it is sensible to extend $\mu$ to $\R^+\cup \{0\}$ with $\mu(p_1,\ldots, p_n) = 0$ when any $p_i = 0$. Moreover, as $\mu$ is continuous as a function of $\tau$, varies strictly monotonically by lemma \ref{AlternatingDerivatives}, and is bounded at infinity by lemma \ref{LEMMA_MuAtInfinity}, we must have that $|\mu(p_1,\ldots, p_{n-1}, \tau)| \in [0,|\mu(p_1,\ldots, p_{n-1})|)$.
\end{proof}

\subsection*{Proof of proposition \ref{PROPOSITION_UniquenessOfMeasure}}
\begin{proof}
We rely on the original result of Baez et al. in \cite{baez2011characterization} which characterises entropy $H$ using conditions on the loss $L$. It therefore suffices to show that entropy loss $L$ and the measure $\mu$ completely determine each other, and that these properties for $\mu$ imply the same properties in $L$ (as the result is stated in \cite{baez2011characterization}).

Firstly, note that setting
\begin{equation}
P = \sum_{i = 1}^n p_i,
\end{equation}
we have, in the spirit of equation \ref{EQUATION_EntropyIsLoss} and due to the homogeneity of $L_d$ that
\begin{align}
L_d\left(\frac{p_1}{P},\ldots,\frac{p_n}{P}\right) & = H_d\left(\frac{p_1}{P},\ldots, \frac{p_n}{P}\right)\\
L_d \, (p_1, \ldots, p_n) & = P^d \cdot H_d \left( \frac{p_1}{P}, \ldots, \frac{p_n}{P} \right).
\end{align}
We note that the Tsallis entropies are also zero on the trivial variable, so we do not need to subtract $H_d(1)$. Using this, and writing $P(\bm{S}) = \sum_{p_i \in \bm{S}} p_i$, coupled with the formula for computing the M\"obius inversion in terms of loss, we have that
\begin{align}
H_d(p_1,\ldots, p_n) & = \sum_{\substack{\bm{S} \subseteq \{p_1,\ldots, p_n\} \\ |\bm{S}|\geq 2}} \mu_d(\bm{S}) \\
\mu_d(p_1,\ldots, p_n) & = \sum_{\substack{\bm{S} \subseteq \{p_1,\ldots, p_n\} \\ |\bm{S}| \geq 2}} (-1)^{n-|S|} P(\bm{S})^d H_d\left(\frac{\bm{S}}{P(\bm{S})}\right).
\end{align}
From which it is now clear that $\mu$, $L$ and $H$ explicitly depend on each other.

The original theorem of Baez, Fritz and Leinster's \cite{baez2011characterization} states that, given a map sending morphisms in the category of finite measure spaces $\cat{FinMeas}$ to numbers in $[0,\infty)$ satisfying functoriality, additivity, homogeneity of degree $d$, and continuity, that this map must be $F(f) = c(H_d(p) - H_d(q))$\cite{baez2011characterization}.

In particular, the measure $L$, which we have now seen is equivalent to specifying $\mu$, is the loss measure specified on morphisms in $\cat{FinMeas}$. By the additive nature of $\mu$ and $L$, homogeneity of $\mu$ is easily seen to be equivalent to the homogeneity of $L$, and continuity of $\mu$ is also equivalent to the continuity of $L$. We therefore need only to demonstrate that if $\mu_d$ satisfies the additivity and functoriality properties, then so too must the loss $L_d$. Applying the result of Baez et al. then shows this is sufficient to characterise $L_d$ and hence $\mu_d$.

Given two independent systems it is straightforward to see that $\mu$, as a measure, should be taken to be additive. Given two morphisms between two pairs of variables $X_1 \to X_2$ and $Y_1 \to Y_2$, each morphism corresponds to a loss $S_X = \Delta X_1 \setminus \Delta X_2$ and $S_Y = \Delta Y_1 \setminus \Delta Y_2$. If $\mu$ is additive so that for any two sets $S_X$ and $S_Y$ $\mu(S_X \sqcup S_Y) = \mu(S_X) + \mu(S_Y)$, then in this can be expressed as a loss $L(X_1 \to X_2) + L(Y_1 \to Y_2)$, so the loss is also additive across independent systems. So additivity of $\mu$ must give additivity of $L$.

For functoriality, we suppose that $\mu$ is functorial in that it is additive down a chain of sets $S_1 \supseteq S_2 \supseteq S_3$ with $\mu(S_1 \setminus S_3) = \mu(S_1 \setminus S_2) + \mu(S_2 \setminus S_3)$. Then given three sets $\Delta X_1 \supseteq \Delta X_2 \supseteq \Delta X_3$ representing a two-step entropy loss, we see that the measure $\mu(\Delta X_1 \setminus \Delta X_3) = \mu((\Delta X_1 \setminus \Delta X_2) \cup (\Delta X_2 \setminus \Delta X_3)) = \mu(\Delta X_1 \setminus \Delta X_2) + \mu(\Delta X_2 \setminus \Delta X_3)$. These quantities then correspond to $L(X_1 \to X_3) = L(X_1 \to X_2) + L(X_2 \to X_3)$, so $L$ must also be functorial.

Hence the functoriality of $\mu$ forces the loss $L$ to also be functorial. Hence $L$ must be uniquely constructed as the loss in $H_d$ up to a scale factor by the result of Baez et al. \cite{baez2011characterization}, which also determines $\mu$.
\end{proof}

\subsection*{Proof of theorem \ref{THM_yeung_correspondence}}
We first state a small lemma which is a standard property of entropy. We will make use of it to demonstrate that our measure is consistent with Yeung's $I$-measure.

\begin{lemma}
\label{LEMMA_partition_law}
Let $P_1,\ldots, P_k$ be disjoint subsets forming a partition of $\Omega$ consisting of individual outcomes $\omega$ of probability $p_\omega$. Then
\begin{equation}
L\left( \sum_{\omega \in P_1} p_\omega, \ldots, \sum_{\omega \in P_k} p_\omega  \right) = L(\Omega) - \sum_{i = 1}^k L(P_i).
\end{equation}
In particular, the expression of the left-hand side is equal to the measure of the subset $\Delta\Omega \setminus \left( \bigcup_{i=1}^k B(P_k) \right).$ 
\end{lemma}
\begin{proof}
We first demonstrate the simple identity
\begin{equation}
L(p_1+p_2,p_3,\ldots, p_n) = L(p_1,p_2,\ldots, p_n) - L(p_1, p_2).
\end{equation}
Let $\Omega = \{\omega_1,\ldots, \omega_N\}$. Then let $X$ be the random variable with partition $\{\{\omega_1,\omega_2\}, \{\omega_3\},\ldots, \{\omega_N\}\}$. By definition we have
\enlargethispage{-3cm} 
\begin{multline}
L(p_1, p_2) = H(\Omega)-H(X) \\
= L(p_1,\ldots, p_n) - L(p_1+p_2,\ldots, p_n),
\end{multline}
giving the identity. The full result then follows by symmetry on the arguments of $L$ and an inductive argument, sequentially decomposing sums into pairs.
\end{proof}

This result essentially states that the total loss of a certain variable defined by the partition $\{P_1,\ldots, P_k\}$ can be computed by calculating the total loss of the entire outcome space and subtracting boundaries internal to parts $P_i$.

We now proceed with the proof of the theorem.

\begin{proof}
We will show that our definition of content agrees with i.) the entropy of individual variables and ii.) the mutual information between two variables. The case for $n$ variables follows inductively.

We will now show that for a variable $X$ with an event space with associated probabilities $p_1,\ldots, p_n$, that $H(X) = L(p_1,\ldots, p_n) = \mu(\Delta X)$, the measure of the content in $X$ (see equation \eqref{EQUATION_EntropyIsLoss}).

Inside of a possibly more refined partition given by outcomes in $\Omega$, we can compute the entropy of $X$ by treating it as a partition $P_1,\ldots, P_k$ of the entire outcome space. In this case it is equivalent to the expression in lemma \ref{LEMMA_partition_law}. As mentioned after the lemma, this corresponds to the measure of the set
\begin{equation}
\Delta\Omega \, \setminus \, \left( \, \bigcup_{i=1}^k \, \left\{b_S: S \subseteq P_i\right\} \right) = \Delta X.
\end{equation}
It can be seen that this is equivalent to the construction of $\Delta X$ in definition \ref{DEFINITION_content}, as the only elements remaining in $\Delta\Omega$ must contain outcomes spanning across partitions. This completes i.).

The mutual information between two variables $X,Y$ is given by
\begin{equation}
\label{Eqn_standard_mutual_information}
I(X;Y) = H(X) + H(Y) - H(X,Y)
\end{equation}
We have seen that $H(V) = \mu(\Delta V)$ for a random variable $V$ inside of a refined space $\Omega$. Given two partitions $P$ and $Q$ corresponding to $X$ and $Y$ respectively, the collection generated by their intersections, $P_i\cap P_j$, is also a partition of $\Omega$, corresponding to the joint random variable $(X,Y)$. This is a refinement of the partitions of $X$ and $Y$.

In particular we have that $b\in \Delta X$ implies $b\in \Delta XY$. Constructing a formal sum of elements $b\in \Delta XY$, we can extend the measure $\mu$ onto this formal sum to obtain
\begin{equation}
I(X;Y) = \mu( \Delta X + \Delta Y - \Delta XY ) = \mu(I)
\end{equation}
Where the formal sum $I = \Delta X + \Delta Y - \Delta XY$ will reflect the mutual information. We see that an atom $b\in \Delta XY$ does not appear in the formal sum $I$ unless $b \in \Delta X \cap \Delta Y$, in which case it appears with coefficient $1$. As all terms in the formal sum have coefficient $1$ or $0$, this formal sum also corresponds to the set of atoms in $\Delta X \cap \Delta Y$. Hence
\begin{equation}
I(X;Y) = \mu(\Delta X \cap \Delta Y).
\end{equation}
That is, our logarithmic decomposition is consistent with standard Shannon mutual information and, by extension, all higher co-informations. It is hence a refinement of the $I$-measure of Yeung \cite{yeung1991new}.
\end{proof}

\subsection*{Proof of theorem \ref{THM_gacs_korner}}
\begin{proof}
The common information variable $Z$ is unique up to isomorphism, so it suffices to demonstrate that this variable $Z$ has its content $\Delta Z \subseteq \bigcap_i \Delta X_i$.

Given an outcome $\omega \in \Omega$, let $\omega$ be contained in the event $X_i(\omega)$ in $X_i$. That is, $\omega$ is contained in one of the parts $X_i(\omega)$ in the partition of $X_i$. By virtue of the definition of the common information, we must have
\begin{equation}
\label{EQN_function_requirements}
f_i(X_i(\omega)) = f_j(X_j(\omega)) \text{ for all $i,j\in\{1,\ldots, n\}$}.
\end{equation}
We will now show the result in two steps. Firstly we show that the common information variable induces a content in $\Delta\Omega$. Then we show that this is contained in the intersection $C$.

Viewing the random variables as partitions of $\Omega$ and using the ordering $A \leq B$ if $A$ is coarser than $B$, we obtain a lattice. Using the restriction in equation \eqref{EQN_function_requirements}, we can see that to compute the partition of $Z$ we must take the meet $X_1\land \cdots \land X_r$ of all variable partitions $X_i$ in the lattice. In particular, the partition of $Z$ has the property that $Z \leq \Omega$, and hence $\Delta Z \subseteq \Delta\Omega$, that is, we have the atoms needed to describe $Z$ in $\Delta\Omega$. Note that $\Delta Z$ might be empty, in which case it corresponds to the trivial random variable.

To show that $\Delta Z$ is contained in the intersection $C = \bigcap_{i} \Delta X_i$, let $b_S \in \Delta Z$. By definition, $S$ crosses a boundary in $Z$. As $Z$ is the finest partition which is coarser than $X_1,\ldots, X_r$, $S$ must cross a boundary in all $X_i$. That is, $b_S \in \bigcap_i \Delta X_i$. Hence $\Delta Z \subseteq C$.

Note that as the partition of $Z$ is unique, the content is also necessarily unique, giving the result.
\end{proof}

\subsection*{Proof of Proposition \ref{PROPOSITION_functional_common_info}}
\begin{proof}
Since $V$ is a deterministic function of the $X_i$, $V$ can be defined by its value for each $\omega \in \Omega$, where $\Omega$ is necessarily finer than the joint outcome space of the $X_i$. As a result, $V$ corresponds to a partition of $\Omega$ and $\Delta V \subseteq \Delta\Omega$, so the functional common information is logarithmically decomposable.

To see that the functional common information is \textbf{not} lattice decomposable, note that selecting the relevant partition often requires reference to the underlying probabilities (e.g. example \ref{EXAMPLE_FunctionalCommonInformation}).
\end{proof}

\subsection*{Proof of Proposition \ref{PROPOSITION_minimum_sufficient_stat}}
\begin{proof}
A statistic $T(X_1,\ldots, X_n) = T(\bm{X})$ is sufficient for some parameter $\theta$ if, for any prior distribution on $\theta$, 
\begin{equation}
    I(T(\bm{X}); \theta) = I(X; \theta).
\end{equation}
In particular, we note that the sample $\bm{X}$ itself is a sufficient statistic. As a result, any minimally sufficient statistic $T$ must have $T = f(\bm{X})$ for a deterministic function $f(\bm{X})$, by definition of minimality. Consequently, the minimum sufficient statistic $T$ can be defined as a partition of $\Omega$, and $\Delta T\subseteq \Delta\Omega$, so that the minimally sufficient statistic is logarithmically decomposable.

However, when finding a minimally sufficient statistic, one has to choose a statistical family from which to model the probabilities. In the case of even four outcomes, these four outcomes could require three free parameters to specify, or only one success parameter if they represent the results of a binomial distribution. That is to say, in order to specify the MSS, one requires data \textit{beyond} the structure of the outcomes alone, which means that the MSS is not lattice decomposable.
\end{proof}

\subsection*{Proof of proposition \ref{PROPOSITION_EntropyExpressions}}
\begin{proof}
Clearly for every entropy expression there is an element of $\Z\Delta\Omega$ (as we can simply find the corresponding entropy contents). We need to check that this representation is unique, and that for any expression $\Z\Delta\Omega$ there is a unique entropy expression.

Suppose that an entropy expression $h$ has two representations $Z_1$ and $Z_2 \in \Z\Delta\Omega$. Since they correspond to the same entropy expression, we must have $\mu(Z_1 - Z_2) \equiv 0$ for all underlying probability distributions. That is, given expressions
\begin{equation}
Z_1 = \sum_{b \in \Delta\Omega} p_b b, \quad Z_2 = \sum_{b \in \Delta \Omega} q_b b
\end{equation}
where $p_b$ and $q_b \in \Z$, we know that
\begin{equation}
\mu\left(\sum_{b \in \Delta\Omega} p_b b\right) \equiv \mu\left(\sum_{b \in \Delta \Omega} q_b b\right).
\end{equation}
As $\mu$ is additive, we can rewrite this as
\begin{equation}
\sum_{b\in \Delta\Omega} (p_b - q_b) \mu(b) = 0.
\end{equation}
We proceed by induction on atom degree. Let $\deg(b) = 2$. Let $\omega_1, \omega_2 \in \Omega$ be any two outcomes. By setting the probability of all outcomes $\omega \in (\Omega \setminus \{\omega_1, \omega_2\})$ to zero, and the probabilities of $\omega_1$ and $\omega_2$ to be both one half, we see that all atoms besides the $\omega_1\omega_2$ atom now have zero measure by lemma \ref{InteriorLossAt0}. Simplifying the sum, we have that
\begin{equation}
(p_{\omega_1 \omega_2} - q_{\omega_1 \omega_2}) \mu(\omega_1, \omega_2) = 0.
\end{equation}
By theorem \ref{AlternatingDerivatives}, we know that $\mu(\omega_1, \omega_2)$ is certainly nonzero, so we have $p_b - q_b = 0$. That is, restricted to all atoms of degree two, the expressions $Z_1$ and $Z_2$ have the same coefficients in $\Z\Delta\Omega$.

We now suppose that all of the coefficients up to degree $d-1$ are equal in $Z_1$ and $Z_2$. By localising in the same fashion to any degree $d$ atom $\omega_1\ldots \omega_d$, we obtain a sum
\begin{equation}
\sum_{\substack{b \in \Delta \Omega \\ b \preceq \omega_1\ldots\omega_d}} (p_b - q_b) \mu(b) = 0.
\end{equation}
However, when performing this `localisation' procedure we are only left with one degree $d$ atom; namely $\omega_1\ldots \omega_d$. So this expression becomes:
\begin{multline}
(p_{\omega_1\ldots \omega_d} - q_{\omega_1\ldots \omega_d}) \mu(\omega_1,\ldots, \omega_d) \\ + \sum_{\substack{b \in \Delta\Omega \\ b \prec \omega_1\ldots \omega_d}} (p_b - q_b) \mu(b) = 0.
\end{multline}
However, by assumption, the entire second sum is precisely zero, yielding $p_{\omega_1\ldots \omega_d} = q_{\omega_1\ldots \omega_d}$. Thus any representation of an entropy expression $h$ is unique in $\Z \Delta \Omega$.

We now need to justify that each element $Z \in \Z\Delta\Omega$ has a corresponding entropy expression. It suffices to show that all single atoms $b \in \Z\Delta\Omega$ have such an expression, from which we can additively derive the entropy expressions of all expressions in $\Z\Delta\Omega$. By considering equation \ref{InclusionExclusion}, we see that all expressions $L(R)$, $R\subseteq \Omega$, by definition, are entropy expressions on $\Omega$. Hence, given some $S\subseteq \Omega$, we have that $\mu(S)$ is an alternating sum of entropy expressions on $\Omega$, giving the result.
\end{proof}

\subsection*{Proof of Theorem \ref{THEOREM_ThreeOperators}}
\begin{proof}
It suffices to prove the three operators commute pairwise.
\begin{itemize}
\item{$\Delta \cdot$ and $\phi(\cdot)$. We consider a single outcome refinement; the rest of the argument follows by extension. Suppose that $\phi: \omega \mapsto \{\omega_1, \omega_2\}$ and that $\Omega = b \cup \{\omega\}$ for some outcomes $\omega_i \in b$, and we have a partition $P = \{b, \{\omega\}\}$. We have that
\begin{equation}
\phi(\Delta P) = \phi(b \omega) = b \omega_1 + b \omega_2 + b \omega_1 \omega_2 = \Delta \phi(P).
\end{equation}
This is sufficient to derive all atoms in $\Delta \phi(P)$. As the refinement is into finitely many parts, we can take every atom in turn and partition successively in two, adding the result each time. If an atom $b\omega$ crosses a boundary in $P$, we know the atoms $b\omega_1, b\omega_2$ and $b\omega_1\omega_2$ cross a boundary in $P'$. These atoms are not provided by any other atom prior to refinement, so this procedure will account for all atoms in $\Delta P'$.}
\item{$\Delta \cdot$ and $\cdot |_S$. Consider an atom $b$ crossing a boundary in $P$ but not completely contained in $S$. Taking $\Delta P$ and restricting to $S$ will eliminate this atom by definition. Similarly, if we restrict to $S$ and consider boundary changes in $S$ only, we will not obtain any atoms not completely contained inside of $S$, so we need only consider atoms contained in $S$. Suppose $b \subseteq S$ is an atom straddling a boundary in $P$. Then $b \in (\Delta P)|_S$ as it is not eliminated when passing to $S$. Similarly, $b$ crosses a boundary in $S$, so $b \in \Delta (P|_S)$. That is, the two sets contain identical atoms.
}
\item{$\phi(\cdot)$ and $\cdot|_S$. We have that
\begin{equation}
\phi(P)|_S = P'|_{S'} = \phi(P|_S).
\end{equation}
Alternatively, the sets $\{a\}, \{b\} \in S$ which are subsets of distinct parts of $P$ lie in distinct parts in $P'$, and hence lie in distinct parts in $S'$.
}
\end{itemize}
\end{proof}

\subsection*{Proof of proposition \ref{PROPOSITION_KullbackLeibler}}
\begin{proof}
This proof is straightforward as it reduces to the limiting density of discrete points of Jaynes \cite{jaynes1957information, jaynes1968prior}. Since we choose the partition of the space $\Omega$ carefully so that the second distribution is uniform, we have, given a discrete variable $X$, that
\begin{align}
D_{\mathrm{KL}}(P(x) \, || \, U(x)) &= \sum_x P(x) \log \frac{P(x)}{U(x)} \notag \\
&= \sum_x P(x) \log P(x) + \sum_x P(x) \log n \notag \\
&= -H(X) + \log n\\
&= -H(X) + H(U) \notag \\
&= -\mu(\Delta X) + \mu(\Delta U), \notag 
\end{align}
as required.
\end{proof}
\ifSubfilesClassLoaded{\bibliographystyle{plain} \bibliography{main} }{}

\begin{thebibliography}{10}

\bibitem{angelov2021explainable}
Plamen~P Angelov, Eduardo~A Soares, Richard Jiang, Nicholas~I Arnold, and
  Peter~M Atkinson.
\newblock Explainable artificial intelligence: {An} analytical review.
\newblock {\em Wiley Interdisciplinary Reviews: Data Mining and Knowledge
  Discovery}, 11(5):e1424, 2021.

\bibitem{baez2011characterization}
John~C Baez, Tobias Fritz, and Tom Leinster.
\newblock A characterization of entropy in terms of information loss.
\newblock {\em Entropy}, 13(11):1945--1957, 2011.

\bibitem{barrett2015exploration}
Adam~B Barrett.
\newblock Exploration of synergistic and redundant information sharing in
  static and dynamical gaussian systems.
\newblock {\em Physical Review E}, 91(5):052802, 2015.

\bibitem{baudot2015homological}
Pierre Baudot and Daniel Bennequin.
\newblock The homological nature of entropy.
\newblock {\em Entropy}, 17(5):3253--3318, 2015.

\bibitem{bell2003co}
Anthony~J Bell.
\newblock The co-information lattice.
\newblock In {\em Proceedings of the Fifth International Workshop on
  Independent Component Analysis and Blind Signal Separation: ICA}, volume
  2003, 2003.

\bibitem{bertschinger2014quantifying}
Nils Bertschinger, Johannes Rauh, Eckehard Olbrich, J{\"u}rgen Jost, and Nihat
  Ay.
\newblock Quantifying unique information.
\newblock {\em Entropy}, 16(4):2161--2183, 2014.

\bibitem{burkart2021survey}
Nadia Burkart and Marco~F Huber.
\newblock A survey on the explainability of supervised machine learning.
\newblock {\em Journal of Artificial Intelligence Research}, 70:245--317, 2021.

\bibitem{campbell1965entropy}
L~Campbell.
\newblock Entropy as a measure.
\newblock {\em IEEE Transactions on Information Theory}, 11(1):112--114, 1965.

\bibitem{down2023logarithmic}
Keenan J.~A. Down and Pedro A.~M. Mediano.
\newblock A logarithmic decomposition for information.
\newblock In {\em 2023 IEEE International Symposium on Information Theory
  (ISIT)}, pages 150--155, 2023.

\bibitem{down2025algebraic}
Keenan~JA Down and Pedro~AM Mediano.
\newblock Algebraic representations of entropy and fixed-sign information
  quantities.
\newblock {\em Entropy}, 27(2):151, 2025.

\bibitem{finn2018pointwise}
Conor Finn and Joseph~T Lizier.
\newblock Pointwise partial information decompositionusing the specificity and
  ambiguity lattices.
\newblock {\em Entropy}, 20(4):297, 2018.

\bibitem{fisher1922mathematical}
Ronald~A Fisher.
\newblock On the mathematical foundations of theoretical statistics.
\newblock {\em Philosophical transactions of the Royal Society of London.
  Series A, containing papers of a mathematical or physical character},
  222(594-604):309--368, 1922.

\bibitem{gacs1973common}
Peter G{\'a}cs and J{\'a}nos K{\"o}rner.
\newblock Common information is far less than mutual information.
\newblock {\em Problems of Control and Information Theory}, 2(2):149--162,
  1973.

\bibitem{gelens2024distributed}
Frank Gelens, Juho {\"A}ij{\"a}l{\"a}, Louis Roberts, Misako Komatsu, Cem Uran,
  Michael~A Jensen, Kai~J Miller, Robin~AA Ince, Max Garagnani, Martin Vinck,
  et~al.
\newblock Distributed representations of prediction error signals across the
  cortical hierarchy are synergistic.
\newblock {\em Nature Communications}, 15(1):3941, 2024.

\bibitem{griffith2014intersection}
Virgil Griffith, Edwin~KP Chong, Ryan~G James, Christopher~J Ellison, and
  James~P Crutchfield.
\newblock Intersection information based on common randomness.
\newblock {\em Entropy}, 16(4):1985--2000, 2014.

\bibitem{griffith2015quantifying}
Virgil Griffith and Tracey Ho.
\newblock Quantifying redundant information in predicting a target random
  variable.
\newblock {\em Entropy}, 17(7):4644--4653, 2015.

\bibitem{griffith2014quantifying}
Virgil Griffith and Christof Koch.
\newblock Quantifying synergistic mutual information.
\newblock In {\em Guided self-organization: inception}, pages 159--190.
  Springer, 2014.

\bibitem{harder2013bivariate}
Malte Harder, Christoph Salge, and Daniel Polani.
\newblock Bivariate measure of redundant information.
\newblock {\em Physical Review E—Statistical, Nonlinear, and Soft Matter
  Physics}, 87(1):012130, 2013.

\bibitem{havrda1967quantification}
Jan Havrda and Franti{\v{s}}ek Charv{\'a}t.
\newblock Quantification method of classification processes. {Concept} of
  structural $ a $-entropy.
\newblock {\em Kybernetika}, 3(1):30--35, 1967.

\bibitem{ince2017measuring}
Robin~AA Ince.
\newblock Measuring multivariate redundant information with pointwise common
  change in surprisal.
\newblock {\em Entropy}, 19(7):318, 2017.

\bibitem{ince2017statistical}
Robin~AA Ince, Bruno~L Giordano, Christoph Kayser, Guillaume~A Rousselet,
  Joachim Gross, and Philippe~G Schyns.
\newblock A statistical framework for neuroimaging data analysis based on
  mutual information estimated via a gaussian copula.
\newblock {\em Human Brain Mapping}, 38(3):1541--1573, 2017.

\bibitem{james2017multivariate}
Ryan~G James and James~P Crutchfield.
\newblock Multivariate dependence beyond {Shannon} information.
\newblock {\em Entropy}, 19(10):531, 2017.

\bibitem{james2018unique}
Ryan~G James, Jeffrey Emenheiser, and James~P Crutchfield.
\newblock Unique information via dependency constraints.
\newblock {\em Journal of Physics A: Mathematical and Theoretical},
  52(1):014002, 2018.

\bibitem{jaynes1957information}
Edwin~T Jaynes.
\newblock Information theory and statistical mechanics.
\newblock {\em Physical Review}, 106(4):620, 1957.

\bibitem{jaynes1968prior}
Edwin~T Jaynes.
\newblock Prior probabilities.
\newblock {\em IEEE Transactions on Systems Science and Cybernetics},
  4(3):227--241, 1968.

\bibitem{kolchinsky2022novel}
Artemy Kolchinsky.
\newblock A novel approach to the partial information decomposition.
\newblock {\em Entropy}, 24(3):403, 2022.

\bibitem{kullback1951information}
Solomon Kullback and Richard~A Leibler.
\newblock On information and sufficiency.
\newblock {\em The Annals of Mathematical Statistics}, 22(1):79--86, 1951.

\bibitem{lehmann2011completeness}
Erich~Leo Lehmann and Henry Scheff{\'e}.
\newblock Completeness, similar regions, and unbiased estimation-part i.
\newblock In {\em Selected works of EL Lehmann}, pages 233--268. Springer,
  2011.

\bibitem{li2023poisson}
Cheuk~Ting Li.
\newblock A poisson decomposition for information and the information-event
  diagram.
\newblock {\em arXiv preprint arXiv:2307.07506}, 2023.

\bibitem{luppi2024synergistic}
Andrea~I Luppi, Pedro~AM Mediano, Fernando~E Rosas, Judith Allanson, John
  Pickard, Robin~L Carhart-Harris, Guy~B Williams, Michael~M Craig, Paola
  Finoia, Adrian~M Owen, et~al.
\newblock A synergistic workspace for human consciousness revealed by
  integrated information decomposition.
\newblock {\em Elife}, 12:RP88173, 2024.

\bibitem{luppi2022synergistic}
Andrea~I Luppi, Pedro~AM Mediano, Fernando~E Rosas, Negin Holland, Tim~D Fryer,
  John~T O’Brien, James~B Rowe, David~K Menon, Daniel Bor, and Emmanuel~A
  Stamatakis.
\newblock A synergistic core for human brain evolution and cognition.
\newblock {\em Nature Neuroscience}, 25(6):771--782, 2022.

\bibitem{marinazzo2022information}
Daniele Marinazzo, Jan Van~Roozendaal, Fernando~E Rosas, Massimo Stella, Renzo
  Comolatti, Nigel Colenbier, Sebastiano Stramaglia, and Yves Rosseel.
\newblock An information-theoretic approach to hypergraph psychometrics.
\newblock {\em arXiv preprint arXiv:2205.01035}, 2022.

\bibitem{mcgill1954multivariate}
William McGill.
\newblock Multivariate information transmission.
\newblock {\em Transactions of the IRE Professional Group on Information
  Theory}, 4(4):93--111, 1954.

\bibitem{mediano2021towards}
Pedro~AM Mediano, Fernando~E Rosas, Andrea~I Luppi, Robin~L Carhart-Harris,
  Daniel Bor, Anil~K Seth, and Adam~B Barrett.
\newblock Towards an extended taxonomy of information dynamics via integrated
  information decomposition.
\newblock {\em arXiv preprint arXiv:2109.13186}, 2021.

\bibitem{mehrabi2021survey}
Ninareh Mehrabi, Fred Morstatter, Nripsuta Saxena, Kristina Lerman, and Aram
  Galstyan.
\newblock A survey on bias and fairness in machine learning.
\newblock {\em ACM Computing Surveys (CSUR)}, 54(6):1--35, 2021.

\bibitem{patil1982diversity}
GP~Patil and Charles Taillie.
\newblock Diversity as a concept and its measurement.
\newblock {\em Journal of the American Statistical Association},
  77(379):548--561, 1982.

\bibitem{proca2024synergistic}
Alexandra~M Proca, Fernando~E Rosas, Andrea~I Luppi, Daniel Bor, Matthew
  Crosby, and Pedro~AM Mediano.
\newblock Synergistic information supports modality integration and flexible
  learning in neural networks solving multiple tasks.
\newblock {\em PLOS Computational Biology}, 20(6):e1012178, 2024.

\bibitem{quax2017quantifying}
Rick Quax, Omri Har-Shemesh, and Peter~MA Sloot.
\newblock Quantifying synergistic information using intermediate stochastic
  variables.
\newblock {\em Entropy}, 19(2):85, 2017.

\bibitem{rosas2019quantifying}
Fernando~E Rosas, Pedro~AM Mediano, Michael Gastpar, and Henrik~J Jensen.
\newblock Quantifying high-order interdependencies via multivariate extensions
  of the mutual information.
\newblock {\em Physical Review E}, 100(3):032305, 2019.

\bibitem{rosas2020operational}
Fernando~E Rosas, Pedro~AM Mediano, Borzoo Rassouli, and Adam~B Barrett.
\newblock An operational information decomposition via synergistic disclosure.
\newblock {\em Journal of Physics A: Mathematical and Theoretical},
  53(48):485001, 2020.

\bibitem{shannon1948mathematical}
Claude Shannon.
\newblock A mathematical theory of communication.
\newblock {\em The Bell System Technical Journal}, 27(3):379--423, 1948.

\bibitem{shannon1953lattice}
Claude Shannon.
\newblock The lattice theory of information.
\newblock {\em Transactions of the IRE Professional Group on Information
  Theory}, 1(1):105--107, 1953.

\bibitem{stanley1986enumerative}
Richard~P Stanley and Richard~P Stanley.
\newblock {\em What is Enumerative Combinatorics?}
\newblock Springer, 1986.

\bibitem{stramaglia2021quantifying}
Sebastiano Stramaglia, Tomas Scagliarini, Bryan~C Daniels, and Daniele
  Marinazzo.
\newblock Quantifying dynamical high-order interdependencies from the
  {O}-information: {An} application to neural spiking dynamics.
\newblock {\em Frontiers in Physiology}, 11:595736, 2021.

\bibitem{te1978nonnegative}
Han Te~Sun.
\newblock Nonnegative entropy measures of multivariate symmetric correlations.
\newblock {\em Information and Control}, 36:133--156, 1978.

\bibitem{ting1962amount}
Hu~Kuo Ting.
\newblock On the amount of information.
\newblock {\em Theory of Probability \& Its Applications}, 7(4):439--447, 1962.

\bibitem{tsallis1988possible}
Constantino Tsallis.
\newblock Possible generalization of {Boltzmann-Gibbs} statistics.
\newblock {\em Journal of Statistical Physics}, 52:479--487, 1988.

\bibitem{vigneaux2017information}
Juan~Pablo Vigneaux.
\newblock Information structures and their cohomology.
\newblock {\em arXiv preprint arXiv:1709.07807}, 2017.

\bibitem{watanabe1960information}
Satosi Watanabe.
\newblock Information theoretical analysis of multivariate correlation.
\newblock {\em IBM Journal of Research and Development}, 4(1):66--82, 1960.

\bibitem{williams2010nonnegative}
Paul~L Williams and Randall~D Beer.
\newblock Nonnegative decomposition of multivariate information.
\newblock {\em arXiv preprint arXiv:1004.2515}, 2010.

\bibitem{wollstadt2023rigorous}
Patricia Wollstadt, Sebastian Schmitt, and Michael Wibral.
\newblock A rigorous information-theoretic definition of redundancy and
  relevancy in feature selection based on (partial) information decomposition.
\newblock {\em Journal of Machine Learning Research}, 24(131):1--44, 2023.

\bibitem{wyner1975common}
Aaron Wyner.
\newblock The common information of two dependent random variables.
\newblock {\em IEEE Transactions on Information Theory}, 21(2):163--179, 1975.

\bibitem{xu2013wyner}
Ge~Xu, Wei Liu, and Biao Chen.
\newblock Wyner's common information: {Generalizations} and a new lossy source
  coding interpretation.
\newblock {\em arXiv preprint arXiv:1301.2237}, 2013.

\bibitem{yeung1991new}
Raymond~W Yeung.
\newblock A new outlook on {Shannon's} information measures.
\newblock {\em IEEE Transactions on Information Theory}, 37(3):466--474, 1991.

\bibitem{yu2016generalized}
Lei Yu, Houqiang Li, and Chang~Wen Chen.
\newblock Generalized common informations: {Measuring} commonness by the
  conditional maximal correlation.
\newblock {\em arXiv preprint arXiv:1610.09289}, 2016.

\bibitem{zhang2021survey}
Yu~Zhang, Peter Ti{\v{n}}o, Ale{\v{s}} Leonardis, and Ke~Tang.
\newblock A survey on neural network interpretability.
\newblock {\em IEEE Transactions on Emerging Topics in Computational
  Intelligence}, 5(5):726--742, 2021.

\end{thebibliography}
\end{document}

\end{document}